\documentclass[preprint,authoryear,12pt]{elsarticle}
\usepackage[a4paper, margin=2.5cm]{geometry} 
\usepackage[utf8]{inputenc}
\usepackage{graphicx}
\usepackage{tabularx}
\usepackage{amsmath, amssymb}
\usepackage{caption,subcaption}
\usepackage{float}
\usepackage{booktabs}
\usepackage{bm}
\usepackage[colorlinks=true,hypertexnames=false]{hyperref}

\hypersetup{
linkcolor=blue,
filecolor=blue,      
urlcolor=blue,
pdftitle={Over the Stability Space of a Multivariate Time Series},
pdfauthor={Roberto~Vásquez-Martínez}
pdfkeywords={High-dimensional analysis, Stable projections, Latent factor modelling},
citecolor=red
}

\newtheorem{theorem}{Theorem}[section]
\newtheorem{lemma}[theorem]{Lemma}
\newtheorem{proposition}[theorem]{Proposition}
\newdefinition{definition}{Definition}[section]
\newdefinition{remark}{Remark}
\newdefinition{assumption}{Assumption}
\newdefinition{algorithm}{Algorithm}

\newcommand{\abs}[1]{\left| #1 \right|}
\newcommand{\norm}[1]{\lVert #1 \rVert}
\DeclareMathOperator{\argmax}{argmax}
\DeclareMathOperator{\cov}{Cov}

\begin{document}

\begin{frontmatter}
\title{Over the Stability Space of a Multivariate Time Series}

\author[1]{Roberto Vásquez Martínez\corref{cor1}}
\ead{v.roberto@lancaster.ac.uk}

\author[2]{Graciela González Farías\fnref{fn2}}

\author[3]{José Ulises Márquez Urbina\fnref{fn4}}

\author[4]{Francisco Corona\fnref{fn3}}

\cortext[cor1]{Corresponding author.}
\fntext[fn2]{Permanent address: Department of Probability and Statistics, Research Centre in Mathematics, Jalisco S/N, Valenciana, Guanajuato, 36023, GTO, Mexico.}
\fntext[fn4]{Permanent address: Department of Probability and Statistics, Research Centre in Mathematics (Monterrey Unit), Alianza Centro No. 502, Technology Research and Innovation Park (PIIT), Apodaca, N.L. 66629, Mexico.}

\affiliation[1]{organization={STOR-i Centre for Doctoral Training, Lancaster University},
            addressline={Science \& Technology Building}, 
            city={Lancaster},
            postcode={LA1 4YR}, 
            country={United Kingdom}}
            
\affiliation[2]{organization={Institute of Data Science and Artificial Intelligence, University of Navarra},
            addressline={Campus Universitario}, 
            city={Pamplona},
            postcode={31009}, 
            country={Spain}}

\affiliation[3]{organization={School of Economics \& Business, University of Navarra},
            addressline={Campus Universitario}, 
            city={Pamplona},
            postcode={31009}, 
            country={Spain}}            

\affiliation[4]{organization={National Institute of Statistics and Geography (INEGI)},
            addressline={Av. Patriotismo 711}, 
            city={Mexico City},
            postcode={03730}, 
            country={Mexico}}

\begin{abstract}
This paper jointly addresses the challenges of non-stationarity and high dimensionality in analysing multivariate time series. Building on the classical concept of cointegration, we introduce a more flexible notion, called stability space, aimed at capturing stationary components in settings where traditional assumptions may not hold. Based on the dimensionality reduction techniques of Partial Least Squares and Principal Component Analysis, we proposed two non-parametric procedures for estimating such a space and a targeted selection of components that prioritise stationarity. We compare these alternatives with the parametric Johansen procedure when possible. Through simulations and real-data applications, we evaluated the performance of these methodologies across various scenarios, including high-dimensional configurations where regularisation techniques are explored, considering a sparse version of the Principal Component Analysis.
\end{abstract}

\end{frontmatter}
\section*{Acknowledgements}
Graciela González Farías and José Ulises Márquez Urbina acknowledge their sabbatical stay at the University of Navarra.

\section*{Declaration of interests}
The authors declare that they have no known competing financial interests or personal relationships that could have appeared to influence the work reported in this paper.

\clearpage
\section{Introduction}

The study of phenomena or processes exhibiting temporal dependence arises naturally in various fields such as economics, environmental sciences, and epidemiology, for example, in the analysis of Gross Domestic Product per capita \citep{james_developing_2012}, global surface temperature \citep{kaufmann_emissions_2006}, or daily mortality rates \citep{bhaskaran_time_2013}. Although statistical methods serve various purposes in this context, many rely --at least to some extent-- on the assumption that the underlying stochastic process is stationary to establish theoretical and practical properties. This is particularly true in time series analysis. 

Let $\{X_n, n \in \mathbb{Z}\}$ denote a time series in $\mathbb{R}^m$. Generally, each of the $m$ components of the series $X$ corresponds to a variable, and these variables interact with one another. Let $\{X_n, n = 1, \dots, T\}$ represent a random sample from the series $\{X_n, n \in \mathbb{Z}\}$. To illustrate the importance of the stationarity assumption, consider the $h$-step-ahead forecasting problem: inferring the distribution of the variables $X_{T+1}, \dots, X_{T+h}$ based on the observed data. It is well-known that both Dynamic Factor Models and Vector Autoregressive (VAR) models rely on the assumption of stationarity for forecasting purposes; see \cite{stock_forecasting_2002} and \cite{lutkepohl_new_2005}, respectively. Under the stationarity assumption and employing a VAR model, it is possible to derive the asymptotic distribution of these variables, thereby enabling the construction of both point forecasts and prediction intervals. Similarly, certain time series clustering algorithms \citep{maharaj_cluster_2000} and changepoint detection methods \citep{killick_optimal_2012} also presuppose stationarity.

In the absence of stationarity, a different approach is required. In fact, numerous real-world applications involve non-stationary features, as illustrated by analysis of business cycles \citep{hamilton_new_1989}, as well as studies in biomedical signal processing and seismology \citep{adak_time-dependent_1998}. Non-stationarity may arise for various reasons, including deterministic trends, structural changes, seasonal behaviours, or unit roots. In this work, we address non-stationarity arising from unit roots or stochastic trends. The concepts and techniques introduced in this work offer a different approach to this longstanding problem, leveraging dimensionality reduction techniques.

Another key aspect of modern data analysis is the prevalence of phenomena involving many interacting variables. In the time series context, this gives rise to the challenge of modelling high-dimensional multivariate time series. High dimensionality typically refers to situations where the number of variables or series, $m$, is on the order of hundreds or thousands, and often substantially exceeds the length of the series, $T$. For example, \citet{pena_empirical_2019} discusses practical cases in which $m$ is ten times greater than $T$. Moreover, when $m \gg 0$, various estimation challenges emerge. In VAR models, for instance, multicollinearity and parameter instability naturally arise. Consequently, it is of practical interest to represent and decompose the information using a smaller number of uncorrelated latent variables -in other words, to perform dimensionality reduction. A comprehensive summary of this approach in the time series context is provided by \citet{lam_estimation_2011}. The techniques introduced therein enable analysis in high-dimensional settings. A recent development of producing latent variables for a forecasting analysis in a high-dimensional setting is in \cite{fan_sufficient_2017}.

To tackle the simultaneous challenges of non-stationarity and high dimensionality, we introduce a methodology building upon the concept of cointegration, as originally formulated by \citet{engle_co-integration_1987}. In this regard, the methodology for analysing cointegration in multivariate time series proposed by \citet{johansen_estimation_1991} is particularly influential, as it allows the estimation of a basis for the so-called cointegration space. However, one of the main limitations of Johansen’s procedure is that it becomes computationally infeasible in high-dimensional systems. Moreover, it is a parametric approach that relies on a set of formal assumptions which may not be plausible in practice. This motivates the introduction of a new concept, referred to as the \emph{stability space}, which we will discuss throughout this work.

There are alternative approaches to Johansen’s method that also target the cointegration context and offer certain advantages, as they are based on non-parametric dimensionality reduction techniques. In this article, we examine the approaches of \citet{muriel_pls_2012}, based on the Partial Least Squares (PLS) algorithm \citep{wold_estimation_1966}, and \citet{harris_principal_1997}, based on Principal Component Analysis (PCA) \citep{hotelling_analysis_1933}. These methods provide a useful starting point, as they have inspired a wide range of interesting extensions and modifications in the literature. Moreover, we argue that the greater flexibility of these approaches may be advantageous for estimating the less restrictive concept of stability space, which, for example, allows the construction of forecasts that are well-behaved in the long run. Additionally, we explore the possibility of considering sparse Principal Component Analysis (SPCA) \citep{zou_sparse_2006} as a natural approach in high-dimensional settings.  

After applying a dimensionality reduction method, one typically obtains a smaller number of components with which to represent the original information optimally according to some criterion. For example, in PCA, it is common to retain the first two principal components. However, these first two principal components are not guaranteed to yield stationary time series -- as we shall demonstrate in this article, one should often expect the opposite. We therefore propose selecting the first principal components that produce stationary time series, which is more appropriate, given that time series methods are typically applied to these latent variables.

This article is organised as follows: Section \ref{sec:cointegration} discusses the concept of cointegration and how it motivates the introduction of the stability space, which is addressed in Section \ref{sec:stability_space}. Section \ref{sec:methodology} describes the most relevant aspects of the methods considered: Johansen’s approach and those based on PLS and PCA. Section \ref{sec:simulations} presents the theoretical aspects of a novel simulation study designed to evaluate these methodologies and the results of this assessment. This simulation procedure is itself of independent interest, as it enables the generation of time series with a specified cointegration space -- or, more accurately, with a given stability space. Finally, in Section \ref{sec:examples} we apply the discussed methodologies in economic practical examples, considering both low- and high-dimensional cases. The paper ends with a discussion in Section \ref{sec:discussion}. The mathematical proofs and the code associated with this work are in the Appendix.

\section{Cointegration notion}
\label{sec:cointegration}

Consider a time series $\{X_n, n \in \mathbb{Z}\}$ taking values in $\mathbb{R}^m$, such that $\mathbb{E}[\lVert X_n \rVert^2] < \infty$.

\begin{definition}
The difference operator $\Delta$ is defined as

\begin{equation*}
\Delta X_n := X_n - X_{n-1} \quad \text{for all } n \in \mathbb{Z}.
\end{equation*}

The series $\{X_n, n \in \mathbb{Z}\}$ is said to be integrated of order $d \in \mathbb{N}$ if $d$ is the smallest natural number such that the series $\{\Delta^d X_n, n \in \mathbb{Z}\}$ is stationary. We denote this property as $X \sim \textsf{I}(d)$. Here, $\Delta^d$ denotes the application of the difference operator $d$ times.
\end{definition}

The cases $d = 0$ and $d = 1$ (a unit root) are commonly considered. In the case $d = 0$, the process is stationary, whereas for $d = 1$, the process variance diverges, and innovations have a permanent effect on the process.

However, when considering the notion of equilibrium among the variables of a multivariate time series, it is possible to combine these non-stationary behaviours to obtain a stationary process. This is precisely the idea of cointegration among time series, a concept rigorously formalised in \citet{engle_co-integration_1987}.

\begin{definition}
Consider
\begin{equation*}
    X_n = [X_{1,n}, \dots, X_{m,n}]' \quad \text{for all } n \in \mathbb{Z},
\end{equation*}
where $X_{i,n}$ is the $i$-th coordinate of the vector $X_n$.

Let $\tilde{X}_i := \{X_{i,n}, n \in \mathbb{Z}\}$ denote the univariate time series corresponding to the $i$-th coordinate of the process $\{X_n, n \in \mathbb{Z}\}$.

The series $\tilde{X}_1, \dots, \tilde{X}_m$ are said to be cointegrated of order $\textsf{CI}(d, b)$, denoted $X \sim \textsf{CI}(d, b)$, if:
\begin{enumerate}
\item $\tilde{X}_i \sim \textsf{I}(d)$ for all $i = 1, \dots, m$;
\item There exists $\beta \in \mathbb{R}^m$, with $\beta \neq 0$, such that
$ \{Z_n := \beta' X_n, n \in \mathbb{Z}\} \sim \textsf{I}(d - b).$
The vector $\beta$ is referred to as a cointegration vector or relation.
\end{enumerate}
\end{definition}

Under the above definition, cointegration relations are not necessarily unique. If there exist at most $r$ linearly independent cointegration vectors, then $r$ is referred to as the cointegration rank. The vector space spanned by these $r$ vectors is known as the cointegration space.

In the particular case $X \sim \textsf{CI}(1,1)$, if $\beta$ is a cointegration relation, then $Z$ is a stationary process. In general, cointegration relations are those that reduce the order of integration of the series involved.

On the other hand, by the Wold Decomposition Theorem, any stationary process $X$ can be expressed as the sum of two uncorrelated processes, $\{V_n, n \in \mathbb{Z}\}$ and $\{U_n, n \in \mathbb{Z}\}$, as follows:
\begin{equation*}
X_n = V_n + U_n \quad \text{for all } n \in \mathbb{Z},
\end{equation*}
where $V$ is a deterministic process fully determined by its past, and $U$ is a purely non-deterministic process satisfying $\mathbb{E}[U_n] = 0$ for all $n$. In what follows, we refer to $U$ as the stochastic component of the process $X$.

We now introduce several important assumptions that will allow us to analyse different scenarios regarding the sources of non-stationarity. These assumptions are necessary for certain results in the analysis of cointegration. However, as will be discussed later, they may be flexibilized when applying the methodologies presented in \citet{harris_principal_1997,muriel_pls_2012}.

\begin{assumption}
\label{ass:nondeterministic}
By the discussion above, although $X$ may be non-stationary, we focus on series without deterministic components. Specifically, we assume that $X$ is a non-stationary stochastic process with zero mean, i.e., a mean-return process.
\end{assumption}

In practice, such deterministic components can be removed, for example, by using the residuals from a regression of the form:
\begin{equation*}
    X_n = a + b n + U_n.
\end{equation*}

This regression assumes that the deterministic component is linear. The validity of this assumption is discussed in Section 15.2 of \citet{hamilton_time_2020}. Alternative approaches can also be considered, depending on the particular problem under study. Moreover, we also assume that the underlying series does not contain seasonal components.

Under Assumption \ref{ass:nondeterministic}, the series $X$ may be considered to have the initial condition $X_0 = 0$, i.e., $X_n = 0$ almost surely for $n \leq 0$.

The scenario that defines the cointegration context for subsequent analysis is summarised in the following assumption:
\begin{assumption}
\label{ass:cointegration}
We consider Assumption \ref{ass:nondeterministic}, and additionally that $X$ follows a VAR model of order $p$ and satisfies $X \sim \textsf{CI}(1,1)$, with cointegration rank $r$.
\end{assumption}

A key result presented in \citet{engle_co-integration_1987}, which links series satisfying Assumption \ref{ass:cointegration} to VECM models \citep{sargan_wages_1964,davidson_econometric_1978}, is the Granger Representation Theorem \citep{engle_co-integration_1987}. We state below the version presented in \citet{lutkepohl_new_2005}:

\begin{theorem}
\label{thrm:rep_granger}
Let $X \sim \text{VAR}(p)$ such that $X \sim \textsf{CI}(1,1)$ and is a purely non-deterministic process.

Suppose it has a VECM representation:
\begin{equation}
\label{eq:VECM_rep}
\Delta X_n := \alpha \beta' X_{n-1} + \sum_{j=1}^{p-1} \Gamma_j \Delta X_{n-j} + \epsilon_n,
\end{equation}
with $X_n = 0$ for $n < 0$, and where $\epsilon$ is white noise with $\epsilon_n = 0$ for $n \leq 0$.

Consider the polynomial
\begin{equation*}
    C(z) := (1 - z) I_m - \alpha \beta' z - \sum_{j=1}^{p-1} \Gamma_j (1 - z) z^j.
\end{equation*}

Suppose the following conditions hold:
\begin{enumerate}
\item If $\det C(z) = 0$, then $|z| \geq 1$.
\item The number of unit roots is $m - r$.
\item $\alpha, \beta \in \mathbb{R}^{m \times r}$, with $\operatorname{rank}(\alpha) = \operatorname{rank}(\beta) = r$.
\end{enumerate}

Then,

\begin{equation*}
    X_n=\mathfrak{F}\sum_{i=1}^n \epsilon_i+\mathfrak{F}^{\star}(L) \epsilon_n + X_0^{\star},
\end{equation*}
where
\begin{equation*}
    \mathfrak{F} = \beta_{\perp} \left[ \alpha_{\perp}' \left( I_m - \sum_{i=1}^{p-1} \Gamma_i \right) \beta_{\perp} \right]^{-1} \alpha_{\perp}'.
\end{equation*}

It follows that $\operatorname{rank}(\mathfrak{F}) = m - r$, $\mathfrak{F}^{\star}(L) \epsilon_n :=\sum_{j=0}^\infty \mathfrak{F}^{\star}_j \epsilon_{n-j}\sim \textsf{I}(0)$, and that $\beta_{\perp}, \alpha_{\perp} \in \mathbb{R}^{m \times (m - r)}$ are matrices whose columns are orthogonal to the columns of $\beta$ and $\alpha$, respectively.
\end{theorem}

From this result, the following proposition can be derived.
\begin{proposition}
\label{prop:granger_proy}
Let $\{X_n, n \in \mathbb{Z}\}$ be a time series in $\mathbb{R}^m$ satisfying Assumption \ref{ass:cointegration}. The following decomposition holds:
\begin{equation}
\label{eq:granger_proy}
X_n = P_{\beta_{\perp}} \left( \theta(1) \sum_{i=1}^{n} \epsilon_i \right) + P_{\beta} X_n + \eta_n \quad \text{for all } n \geq 0,
\end{equation}
where $P_{\beta}, P_{\beta_{\perp}}$ are projection matrices onto the column spaces $\operatorname{Col}(\beta)$ and $\operatorname{Col}(\beta_{\perp})$, respectively, $\{\eta_n, n \in \mathbb{Z}\}$ is a stationary time series, and:
\begin{equation*}
    Z_n := P_{\beta} X_n + P_{\beta_{\perp}} \Delta X_n,
\end{equation*}
has a moving average (MA) representation
\begin{equation*}
    Z_n = \theta(B) \epsilon_n.
\end{equation*}

Thus, the series $\{ P_{\beta} X_n, n \in \mathbb{Z} \}$ represents the stationary part of the series $\{ X_n, n \in \mathbb{Z} \}$.
\end{proposition}

For completeness, we notice that the assumption that only deviations from stationarity are influenced by short-term movements might not be adequate in several applications; thus, the  \eqref{eq:VECM_rep} result in a misspecification as pointed out in \cite{scheiblecker_between_2013}. An important extension to consider for the cointegration notion to solve this possible misspecification is the concept of multicointegration introduced in \cite{granger_investigation_1989}. To illustrate it, let a $\{X_n=[X_{1,n},X_{2,n}]',n\geq 0\}$ be a $\mathsf{CI}(1,1)$ system with values in $\mathbb{R}^2$. Then, there exists $\beta=[\beta_1,\beta_2]'\in \mathbb{R}^2$ such that if 
\[
Z_n := \beta'X_n \quad \text{ for }n\in \mathbb{Z},
\]
then $\{Z_n, n\in \mathbb{Z}\}\sim \mathsf{I}(0)$. This is the so-called first level of cointegration that we have discussed before. Now, let the partial sum $S_n=\sum_{i=1}^nZ_i$. If the series $\{S_n,n\geq 0\}$ cointegrates with either $\{X_{1,n},n\geq 0\}$ or $\{X_{2,n},n\geq 0\}$ we get another cointegration relationship and we say the system $\{X_n,n\geq 0\}$ is multicointegrated. Without of generality, we can assume $\{S_n,n\geq 0\}$ and $\{X_{1,n},n\geq 0\}$ are cointegrated,  if $\gamma=[\gamma_{1},\gamma_{2}]'$ is that relation then the process defined by
\begin{align*}
W_n&:=\gamma_1 S_n + \gamma_2 X_{1,n}\\
&=\gamma_1\cdot (\beta_1\sum_{i=1}^n X_{1,i}+\beta_2\sum_{i=1}^n X_{2,i})+\gamma_2X_{1,n},
\end{align*}
is a $\mathsf{I}(0)$ process. Hence, we observe the process $\{W_n-\gamma_2X_{1,n},n\geq 0\}\sim \mathsf{I}(1)$ then the vector $\gamma_1\cdot \beta$ is a cointegrated relation of the $\mathsf{CI}(2,1)$ system $[\sum_{i=1}^n X_{1,i},\sum_{i=1}^n X_{2,i}]$. Based on the work in \cite{johansen_representation_1992,johansen_statistical_1995}, \cite{engle_grangers_1999} demonstrates the importance of a system of $\mathsf{I}(2)$ time series for embodying multicointegrated structures in VAR systems \citep{kheifets_fully_2023}. In fact, our main focus in this work is to consider a VAR system, which is widely used in practice. \cite{engle_grangers_1999} proves that under Assumption~\ref{ass:cointegration}, multicointegration can not occur; although, in a $\mathsf{I}(1)$ system but not $\mathsf{CI}(1,1)$, the system can be reformulated as a set of $\textsf{I}(2)$ series where multicointegration can be shown to result in cointegration amongst the generated $\textsf{I}(2)$ and their first differences \citep{engsted_multicointegration_1999}. Therefore, we can primarily focus on the first level of cointegration in VAR systems in the subsequent development of this work, in addition to the fact that it is the primary practical setting. 

\section{Stability Space}
\label{sec:stability_space}
Proposition \ref{prop:granger_proy} motivates the idea of considering a vector space onto which the projection of the time series under study is stationary. In principle, under Assumption \ref{ass:cointegration}, this vector space coincides with the cointegration space. However, without this restriction, such a space may still exist, but it would no longer correspond to the cointegration space, representing instead a different notion.

Firstly, Assumption \ref{ass:nondeterministic}, despite allowing for a non-stationary process, attempts to simplify the problem by considering this non-stationarity to arise solely from a time series without a determinist component. Assumption \ref{ass:cointegration} is more restrictive and is not valid, for instance, for a series $\{X_n,n \in \mathbb{Z}\}$ whose components exhibit different orders of integration, which is the main setting we might find in practice, specifically, we start considering a combination of $\mathsf{I}(0)$, $\mathsf{I}(1)$  and $\mathsf{I}(2)$ that may represent the intermediate scenarios between a $\mathsf{CI}(1,1)$ system and a system of $\mathsf{I}(2)$ series. 

Thus, the notion of a vector $\bm{v} \in \mathbb{R}^m$ such that the series $\{\bm{v}' X_n,n\ge 0\}$ is stationary is not subject to either Assumption \ref{ass:nondeterministic} or Assumption \ref{ass:cointegration}, which motivates the following definition.
\begin{definition}
    Let $\{X_n,n\geq 0\}$ be a time series in $\mathbb{R}^m$. We say that $\bm{v} \in \mathbb{R}^m$ is a stability relation if the time series $\{\bm{v}' X_n,n\ge 0\}$ is stationary. 

    Moreover, we define the stable space as
    \[
    \mathcal{H}^S:=\{\bm{v} \in \mathbb{R}^m\ |\ \bm{v}\text{ is a stability relation}\}.
    \]
    The space $\mathcal{H}^N:=(\mathcal{H}^S)^\perp$ is referred to as the instability space.
\end{definition}

\begin{remark}
\label{rem:cointegration_relation}
    Under Assumption \ref{ass:cointegration}, the subspace $\mathcal{H}^S$ coincides with the cointegration space. Using the notation of Proposition \ref{prop:granger_proy}, we have $\mathcal{H}^S=\text{Col } \beta$.
\end{remark}

\begin{remark}
\label{rem:multicointegration_relaiton}
    In the multicointegration case for a VAR process with $\mathsf{I}(1)$ series, based on the $\mathsf{I}(2)$ VAR specification \citep{johansen_representation_1992,johansen_statistical_1995} and following \cite{engsted_multicointegration_1999}, we can see the stable space coincides with vector space of dimension $r$ spanned by the cointegrated relationships to the $\mathsf{I}(0)$ level, we also have $s$ relationships cointegrated to the so-called $\mathsf{I}(1)$ common trends and $m-r-s$ $\mathsf{I}(2)$ common trends.     
\end{remark}

\begin{remark}
    The notion of the stable space could be applied to a time series without distinguishing the underlying cause of its non-stationarity (unit root, structural break, deterministic trend, fractional order of integration), although this idea would require further formalisation.
\end{remark}

Therefore, unlike Johansen's procedure, which heavily depends on Assumption \ref{ass:cointegration}, this is not the case for the methodologies proposed in \citet{muriel_pls_2012} and \citet{harris_principal_1997}. In fact, \citet{wold_model_1980} notes that the PLS algorithm can be used in complex variable systems with few theoretical assumptions. Thus, these methodologies can be employed to identify a basis for the stable space $\mathcal{H}^S$.

Assuming $\dim \mathcal{H}^S = r$, each of these methodologies yields an estimated matrix $\hat{\beta} = [\hat{\beta}_1, \dots, \hat{\beta}_{\hat{r}}]$, whose columns form an estimated basis of the stable space $\mathcal{H}^S$. Here, $\text{Col } \hat{\beta}$ denotes the estimated space, and $\hat{r}$ is the estimated dimension, which may vary across methods. In principle, $\hat{\beta}$ differs from one method to another, although the vector space $\text{Col } \hat{\beta}$ may coincide. Further details on the estimation procedure are provided in Section~\ref{sec:methodology}.

The column vectors $\hat{\beta}_i$, for $i=1,\dots,\hat{r}$, will represent the estimated stable components of the time series $X$, and
\[
T_i(n):= \hat{\beta}_i'X_n\quad \text{ for }n\geq 0,
\]
the stationary scores associated with these components.

The stationary scores represent the relationships among the variables comprising the time series. For example, let $\hat{\beta}_1=[\hat{\beta}_{1,1},\dots,\hat{\beta}_{1,m}]'$, then, the weights $\hat{\beta}_{1,j}$ indicate the extent and manner in which the variable $X_{j,n}$ contributes to the score $T_1(n)$. In this way, this concept allows one to obtain stationary latent variables from a vector time series by combining information from its constituent series. Thus, these scores reveal how the series within the system compensate each other to achieve long-term stability. These stationary scores might be used for forecasting tasks. In that venue, an idea of the use of stationary scores for prediction is developed in \cite{wang_combined_2021} for wind power series forecasting. However, the authors used a different approach for the stationary scores estimation, and they do not draw any theoretical development as we did here, where we considered its connection with the cointegration/multicointegration notion.

\section{Methodology}
\label{sec:methodology}
We shall now describe the various methodologies that will be considered for estimating the columns of $\beta = [\beta_1, \dots, \beta_r]$, which will correspond to the estimated basis of the stable space $\mathcal{H}^S$. All methodologies begin with a random sample $\{X_n, n = 1, \dots, T\}$ from a time series taking values in $\mathbb{R}^m$. For simplicity, we only consider methodologies initially developed for $\mathsf{CI}(1,1)$ because it is the most common case in user cases and because we want to test later deviations from this kind of process in the way of considering a mixture of integration orders within the variables in the system.

\subsection{Johansen Procedure}
\label{sub:method_johansen}
One of the most widely used methods for estimating the cointegration space is the Johansen procedure, introduced in \citet{johansen_estimation_1991}. This methodology relies heavily on Assumption \ref{ass:cointegration} and is based on the VECM parametrisation of $X$ in \eqref{eq:VECM_rep}.

We now summarise the steps that constitute the Johansen procedure. Let 
\[
M_{p-1}(n):=\textsc{span}\{\Delta X_{n-1},\dots,\Delta X_{n-p+1}\} 
\] 
Using the Projection Theorem, we can obtain the decompositions
\[
  \Delta X_{n}=R_{0n}+P_{M_{p-1}(n)}\Delta X_{n} \in \mathbb{R}^{m},
\]
and
\[
  X_{n-1}=R_{1n}+P_{M_{p-1}(n)}X_{n-1} \in \mathbb{R}^{m},
\]
where $R_{0n}=P_{M_{p-1}(n)^\perp}\Delta X_{n}$ and $R_{1n}=P_{M_{p-1}(n)^\perp}X_{n-1}$.

Therefore, by applying $P_{M_{p-1}(n)^\perp}$ to \eqref{eq:VECM_rep}, we obtain the relation
\[
R_{0n}= \alpha \beta' R_{1n}+\epsilon_n,
\]
with $\{\epsilon_n,n=1,\dots,T\}$ being centred Gaussian noise.

The Johansen procedure estimates $\alpha$ and $\beta$ from the above relation using maximum likelihood, since Gaussian noise is assumed.

The main interest lies in estimating $\beta$. In \citet{maddala_unit_1998}, it is shown that, under normality and assuming the cointegration rank $r$ is known, the maximum likelihood estimator of $\beta$ is equivalent to the matrix whose columns correspond to the first $r$ asymptotic canonical correlations between $R_{0n}$ and $R_{1n}$.

This is due to the fact that we attempt to linearly predict an $\textsf{I}(0)$ process ($\Delta X_n$) with an $\textsf{I}(1)$ process ($X_{n-1}$), and thus, intuitively, we seek the linear combinations of $X_{n-1}$ that are most correlated with $\Delta X_{n}$.

To obtain an estimator $\hat{r}$ of $r$, \citet{johansen_estimation_1991} proposes a method based on the likelihood ratio test. This is described in Section 6.5.1 of \citet{maddala_unit_1998}. In summary, the estimator $\hat{r}$ is obtained by successively applying the likelihood ratio test
\[
H_0: r\leq r_0\quad \text{vs}\quad H_a: r> r_0,
\]
whose test statistic is
\[
\Lambda_{r_0}:=-T\sum_{i=r_0+1}^m \log(1-\hat{\lambda}_i),
\]
where $\hat{\lambda}_i$ corresponds to the $i$-th estimated canonical correlation.

When, for the first time, there is no evidence to reject the null hypothesis as the value of $r_0$ increases, an estimator is obtained—in this case, of the cointegration rank, that is, $\hat{r} = r_0$.

We note that the Johansen procedure is based on the following assumptions:
\begin{enumerate}
    \item Assumption \ref{ass:cointegration}.
    \item A Gaussian innovation process for the underlying process $X$. The Johansen procedure is sensitive to this assumption, as illustrated in \citet{gonzalo_pitfalls_1998,huang_long-run_1996}.
\end{enumerate}

Moreover, \citet{maddala_unit_1998} indicates that critical values for this hypothesis test can be found in \citet{osterwald-lenum_note_1992} for processes with $m \leq 11$. Precisely because of this, implementations of this method, such as the \textsf{ca.jo} class from the \textsf{urca} \textsf{R} package \citep{pfaff_analysis_2008}, only support systems with $m \leq 11$.

Theoretically, it is possible to carry out the Johansen procedure for any value of $m$, however, it becomes computationally prohibitive. For this reason, the methodology is limited in the context of high dimensionality. Thus, the parametric nature of this methodology, Assumption \ref{ass:cointegration}, and the issue of high dimensionality are all constraints that limit the practical applicability of the Johansen procedure.

\subsection{PCA}
\label{sub:method_PCA}
A first alternative for the estimation of $\beta$ is the approach proposed in \citet{harris_principal_1997}. In \citet{harris_principal_1997}, the estimation of $\beta$ is carried out under Assumption \ref{ass:cointegration}; however, we propose to apply the same methodology while relaxing these assumptions.

The main motivation lies in the intuition behind the Johansen procedure, which seeks to linearly predict $\Delta X_n$ using $X_{n-1}$. In order to avoid differencing the series and to perform the analysis at the level of the original series, the core idea of the methodology proposed by \citet{harris_principal_1997} is to analyse the variability of the original series $\{X_n, n=1, \dots, T\}$.

For example, under Assumption \ref{ass:cointegration}, $X_n$ has components that are at least $\textsf{I}(1)$, and we aim to predict $X_n$ by identifying linear combinations of its components that yield a stationary time series. These linear combinations are expected to exhibit lower variability, as they correspond to an $\textsf{I}(0)$ process. Consequently, PCA is based on the notion that the stable linear combinations of a process with non-stationary components should correspond to the principal components with the lowest variability.

Therefore, if $r$ is known, an estimator for the columns of $\beta$ is the matrix formed by the $r$ eigenvectors associated with the $r$ smallest eigenvalues from the spectral decomposition of the sample covariance matrix
\[
S := T^{-1} \sum_{i=1}^{T} X_i X_i'.
\]

In the steps of this methodology, Assumption \ref{ass:cointegration} is not theoretically essential, unlike in the Johansen procedure, although it is necessary for deriving the asymptotic properties of the obtained estimator, as detailed in \citet{harris_principal_1997}.

Thus, to estimate $\beta$ and the dimension $r$ of the stable space $\mathcal{H}^S$, one first computes the eigenvectors $\hat{\beta}_1,\dots,\hat{\beta}_m$ associated with the eigenvalues
\[
\hat{\lambda}_1\geq \dots\geq \hat{\lambda}_m,
\]
of $S$. Subsequently, we consider
\[
T_i(n)=\hat{\beta}_i'X_n\quad \text{ for }n=1,\dots,T,
\]
and  the hypothesis test
\begin{equation}
\label{eq:hypothesis_test_PCA}
H_0:\ r\geq r_0\quad \text{vs}\quad H_a:\ r< r_0.
\end{equation}

Moreover, as noted in \citet{harris_principal_1997}, it is observed that for $i < j$, if the series $\{T_i(n), n=1,\dots,T\}$ is stationary, then $\{T_j(n), n=1,\dots,T\}$ is also stationary.

Therefore, the hypothesis test in \eqref{eq:hypothesis_test_PCA} is equivalent to
\[
H_0:\ T_{m-r_0+1}\sim \textsf{I}(0)\quad \text{vs}\quad H_a:\ T_{m-r_0+1}\sim \textsf{I}(1),
\]
and this test can be carried out, for example, using the KPSS test \citep{kwiatkowski_testing_1992}, while, if the null and alternative hypotheses are switched, the augmented Dickey–Fuller (ADF)\citep{dickey_unit_1986} test may be used. However, it is more convenient to consider stationarity in the null hypothesis—first, because we are logically seeking statistical evidence to rule out non-stationary scores, and second, because it avoids the issue that arises from the presence of multiple unit roots when using the ADF test setup, as discussed in \citep{dickey_determining_1987}. Indeed, the recommended scenario outside the scope of Assumption \ref{ass:cointegration} is to consider different higher orders of integration, as will be detailed in Section \ref{sec:simulations}.

Although $\hat{r}$ is the first value of $r_0$ for which the null hypothesis of the previous test is not rejected—under Assumption \ref{ass:cointegration}—we consider relaxing this assumption. Consequently, we take as the estimator of $\beta$ the components that give rise to stationary scores, that is,
\begin{equation*}
    \hat{\beta}_{\textsc{PCA}}:=[\hat{\beta}_{i},\dots,\hat{\beta}_{i_{\hat{r}}}],
\end{equation*}
where $\hat{\beta}_{i_k}$ for $k=1,\dots,\hat{r}$ denotes the set of components, which are not necessarily consecutive but are ordered by importance with respect to the spectral decomposition, such that $T_{i_k}$ is not rejected as being a stationary series. The subscript indicates the method used for the estimation.

We observe that this PCA-based method relies on the decomposition of the variance of the process $X$, rather than on the dependence between the variables $X_n$ and $X_{n-1}$, which naturally arises in a time series. This aligns with the intuition behind the following methodology. On the other hand, because our interest is also in high-dimensional settings, we consider using sparse PCA (SPCA) introduced in \cite{zou_sparse_2006}, and, to our understanding, it has not been used in this context. The stationary scores detection mechanism remains the same; the intention is only to introduce sparsity in the scores loadings to favour interpretation and might improve the stable space estimation in a high-dimensional configuration. 

\subsection{PLS}
\label{sub:method_PLS}
It is of interest to analyse the covariance between the variables $X_n$ and $X_{n-1}$—which is ignored by the PCA-based methodology but is considered in the Johansen procedure, although the latter does so between $\Delta X_n$ and $X_{n-1}$. Moreover, obtaining stationary scores leads to identifying the directions $\hat{w}_i$ with the lowest variability, as in \citet{harris_principal_1997}. These are the main motivations for the methodology proposed in \citet{muriel_pls_2012}.

In \citet{muriel_pls_2012}, a PLS procedure is proposed to estimate cointegrating vectors, assuming that
\begin{equation}
   \label{eq:natural-choice}
   \mathbf{Y}:=[X'_{T},\dots,X'_2]'\in\mathbb{R}^{(T-1)\times m} \quad \text{ and }\quad \mathbf{X}:=[X'_{T-1},\dots,X'_{1}]' \in \mathbb{R}^{(T-1)\times m},
\end{equation}
represent the response and the predictor variables, respectively.

However, one may also consider a series $\{Y_n, n=1, \dots, T\}$ taking values in $\mathbb{R}^p$ as the target, and the series $\{X_n, n=1, \dots, T\}$ in $\mathbb{R}^m$ as the covariate.

It is also common that $p \ll m$, in which case techniques are required to generate predictions of $Y$ by concentrating the information from $X$ into a smaller number of latent variables so that this does not become a limitation.

In this context, the PLS algorithm naturally arises as an alternative. The PLS algorithm is an iterative process in which the first iteration corresponds to solving the following optimisation problem:
\begin{equation}
  \label{eq:PLS_TimeSeries}
  (\hat{w}_1,\hat{c}_1):=\argmax_{d \in \mathbb{R}^{m},e \in \mathbb{R}^{m}}\cov(\mathbf{X}d,\mathbf{Y}e)^2\quad \text{ with }\lVert d\rVert=\lVert e\rVert=1.
\end{equation}

This optimisation problem is solved at each iteration, but taking into account the information not captured by the components computed up to that point. A simplification of this optimisation problem is described in the following lemma:
\begin{lemma}
\label{lem:simplified_optimisation}
Let $\{X_n, n=1,\dots,T\}$ and $\{Y_n, n=1,\dots,T\}$ be processes taking values in $\mathbb{R}^m$ and $\mathbb{R}^p$, respectively. Let $\mathbf{Y}$ and $\mathbf{X}$ be the corresponding design matrices.

Considering the first iteration of the PLS algorithm in equation \eqref{eq:PLS_TimeSeries}, it holds that
\begin{equation}
\label{eq:Ycomponents}
\hat{c}_1 \propto \mathbf{Y}' \mathbf{X} \hat{w}_1,
\end{equation}
and therefore, the optimisation problem in equation \eqref{eq:PLS_TimeSeries} can be simplified to
\[
\hat{w}_1 := \argmax_{d \in \mathbb{R}^{m}} d' \mathbf{X}' \mathbf{Y} \mathbf{Y}' \mathbf{X} d \quad \text{subject to } \lVert d\rVert=1,
\]
which implies that $\hat{w}_1$ is the eigenvector corresponding to the largest eigenvalue of the matrix $\mathbf{X}' \mathbf{Y} \mathbf{Y}' \mathbf{X}$. Moreover, the proportionality constant in the relation \eqref{eq:Ycomponents} is the reciprocal $\lambda^{-1}$ of the largest singular value of the sample cross-covariance $\mathbf{Y}'\mathbf{X}$.
\end{lemma}

The PLS algorithm in the context of regression between $\mathbf{Y}$ and $\mathbf{X}$ is described below, based on the steps outlined in \citet{hoskuldsson_pls_1988}:

\begin{algorithm}
\label{algo:PLS_TS}
We begin with a centred and scaled random sample from time series $\{X_n,n=1,\dots,T\}$ and $\{Y_n,n=1,\dots,T\}$ as in Lemma \ref{lem:simplified_optimisation}.  

Let $\mathbf{X}^{(1)}$ and $\mathbf{Y}^{(1)}$ be the corresponding design matrices.

For $i = 1, \dots, m$
\begin{enumerate}[Step 1.]
  \item Compute $\hat{w}_i$ according to Lemma \ref{lem:simplified_optimisation}, and calculate the variance:
  \[
  S^{(i)} := T^{-1}(\mathbf{X}^{(i)})'\mathbf{X}^{(i)},
  \]
  corresponding to the information at the current iteration.
  \item Define the variance-weighted projection matrix:
  \[
  P_{i} := I_m - \frac{\hat{w}_i\hat{w}_i'S^{(i)}}{\hat{w}_i' S^{(i)}\hat{w}_i}\quad \text{ where }I_m\text{ is the identity matrix of dimension }m.
  \]
  \item Compute the score $T_i = \mathbf{X}^{(i)}\hat{w}_i$ and the projection matrix:
  \[
  Q_{i} := I_{T-1} - \frac{T_iT_i'}{T_i'T_i}\quad \text{ where }I_{T-1}\text{ is the identity matrix of dimension }T-1.
  \]
  \item If $i < m$, obtain the residual information:
    \[
    \mathbf{X}^{(i+1)} := \mathbf{X}^{(i)}P_i,
    \]
    and
    \[
    \mathbf{Y}^{(i+1)} := Q_i\mathbf{Y}^{(i)}.
    \]
  \item Continue to Step 1.
\end{enumerate}
\end{algorithm}

\begin{remark}
\label{rem:simplified_update}
In each iteration of Algorithm \ref{algo:PLS_TS}, and in the particular case where $\mathbf{X}^{(1)}$ and $\mathbf{Y}^{(1)}$ are taken as in \eqref{eq:natural-choice}, $S^{(i)}$ is an estimator of the variance of the residual information of the series.

Moreover, it can be shown -- based on the algorithm described in \citet{hoskuldsson_pls_1988} -- that the residual information associated with the explanatory variables can also be updated via
\[
\mathbf{X}^{(i+1)} = Q_i\mathbf{X}^{(i)}.
\]

Although Algorithm \ref{algo:PLS_TS} follows the steps formulated, for example, in \citet{hoskuldsson_pls_1988,garthwaite_interpretation_1994}, it can be simplified by updating only $\mathbf{X}$ while keeping the matrix $\mathbf{Y}$ fixed throughout the algorithm, as shown in the \ref{appen:proofs}.
\end{remark}

\begin{remark}
Let $s_i(A)$ denote the $i$-th largest singular value of the matrix $A$. In \citet{hoskuldsson_pls_1988}, it is noted that for $i \geq 1$
\begin{equation}
\label{eq:SVD_inequality}
s_2((\mathbf{X}^{(i)})'\mathbf{Y})^2 \leq s_1((\mathbf{X}^{(i+1)})'\mathbf{Y})^2 \leq s_1((\mathbf{X}^{(i)})'\mathbf{Y})^2,
\end{equation}
Considering that singular values are associated with the amount of information concentrated in their respective components, the inequality above implies that $\hat{w}_{i+1}$ captures more information than the second singular vector. In this sense, the components obtained via the PLS algorithm are more efficient than simply using the singular value decomposition (SVD) decomposition of the cross-covariance matrix.
\end{remark}

The objective is to estimate a basis for the stable space $\mathcal{H}^S \subset \mathbb{R}^m$ associated with the series $\{X_n, n=1, \dots, T\}$. In this way, these components will concentrate the information of the series into components that yield stationary scores. To determine which components $\hat{w}_i$ belong to the stable space and which do not, we proceed similarly to Section \ref{sub:method_PCA}.

It is of interest that these scores be linear combinations of the original information from the series $\{X_n, n=1, \dots, T\}$ for interpretability purposes, as specified in the following proposition:
\begin{proposition}
\label{prop:pesosBeta}
For each $i = 1, \dots, m$, let $A_i \in \mathbb{R}^{m \times m}$ such that
\[
A_i = \begin{cases}
    I_m &\ \text{if } i = 1, \\
    \prod_{k=1}^{i-1} P_k &\ \text{if } i > 1,
\end{cases}
\]
then
\[
\mathbf{X}^{(i)} = \mathbf{X}^{(1)} A_i.
\]
Thus, if we define
\[
\hat{\psi}_i := A_i \hat{w}_i \in \mathbb{R}^{m},
\]
then
\[
T_i = \mathbf{X}^{(i)} \hat{w}_i = \mathbf{X}^{(1)} \hat{\psi}_i.
\]
\end{proposition}

The previous proposition tells us that the weights $\hat{w}_i$ define the score $T_i$ based on the residual information at the $i$-th iteration, while $\hat{\psi}_i$ are the corresponding weights used to construct the score in terms of the original information. Moreover, one of the properties of Algorithm \ref{algo:PLS_TS} is that it yields orthogonal components $\hat{w}_i$ and scores $T_i$, that is, we obtain a decomposition of the original information into uncorrelated scores.

To distinguish which components of $\{\hat{w}_i, i=1,\dots,m\}$ give rise to stationary scores and which do not, we follow the same approach as in Section \ref{sub:method_PCA}, selecting those that yield stationary scores. In this way, the estimator $\hat{r}$ is the number of components that result in stationary scores, and
\[
\hat{\beta}_{\textsc{PLS}} := [\hat{w}_{i_1}, \dots, \hat{w}_{i_{\hat{r}}}],
\]
in the same manner as in the previous section.

In \citet{muriel_pls_2012}, the KPSS test is also used. Along with the PLS or PCA method, the methodology is also identified by the hypothesis test employed; that is, the methodology introduced in \citet{muriel_pls_2012} corresponds to the PLS-KPSS method.

Moreover, \citet{muriel_pls_2012} considers the particular case in which $\mathbf{X}^{(1)}$ and $\mathbf{Y}^{(1)}$ are constructed as in \eqref{eq:natural-choice}. In that case, a basis for the stable space $\mathcal{H}^S$ associated with the series $\{X_n, n=1, \dots, T\}$ is estimated using Algorithm \ref{algo:PLS_TS} and the hypothesis testing scheme.

Under Assumptions \ref{ass:nondeterministic} and \ref{ass:cointegration}, the cointegration space is estimated, and a result similar to Lemma 1 of \citet{harris_principal_1997} on the consistency of the estimator $\hat{\beta}_{\text{PLS}}$ is obtained:
\begin{proposition}
\label{prop:consistenciaPLS}
    Let $X$ be a time series taking values in $\mathbb{R}^{m}$ satisfying Assumption \ref{ass:cointegration} and with an innovation process given by strong white noise. Suppose that $r = \dim \mathcal{H}^S$ is known. 
    
    Let $\{X_n, n=1, \dots, T\}$ be a random sample from which the estimator $\hat{\beta}_{\mathrm{PLS}}$ is obtained by initialising Algorithm \ref{algo:PLS_TS} as in \eqref{eq:natural-choice}.

    Let $\beta$ and $\beta_{\perp}$ be as in Theorem \ref{thrm:rep_granger}. Then it holds that  
    \[
    P_{\beta_{\perp}}\hat{\beta}_{\mathrm{PLS}} = (I_m - \beta(\beta'\beta)^{-1}\beta')\hat{\beta}_{\mathrm{PLS}} = O_p(T^{-1}).
    \]
\end{proposition}

As a final remark, in a dimensionality reduction context, it is necessary to represent the information through such latent variables -- or, in this case, scores. We observe that the PLS algorithm implicitly decomposes the observed time series through the following relation:
\begin{equation}
\label{eq:PLS_decomposition}
\mathbf{Y} = \sum_{j \not\in \{i_1, \dots, i_{\hat{r}}\}} T_j \hat{c}_j' + \underbrace{\sum_{k=1}^{\hat{r}} T_{i_k} \hat{c}_{i_k}'}_{\text{stationary part}},
\end{equation}
therefore, unlike the Johansen procedure and PCA, the regression step is not necessary when aiming to approximate the observed information using the scores of interest -- in this case, the stationary scores.

\section{Simulation study}
\label{sec:simulations}
We compare the estimation of the stable space for a multivariate time series $\{X_n, n=1, \dots, T\}$ under two scenarios. 
\begin{itemize}
    \item \textit{Scenario 1:} Assumptions \ref{ass:nondeterministic} and \ref{ass:cointegration} hold, i.e., the components of $\{X_n, n=1, \dots, T\}$ are cointegrated. 
    \item \textit{Scenario 2:} Assumption \ref{ass:nondeterministic} holds, but the components of $\{X_n, n=1, \dots, T\}$ exhibit different orders of integration. In particular, we consider a mixture of $\textsf{I}(0)$, $\textsf{I}(1)$, and $\textsf{I}(2)$ components, as these are the most commonly encountered orders of integration in practice. In this case, Assumption \ref{ass:cointegration} completely is not satisfied.
\end{itemize}

The idea for constructing simulations of the process $\{X_n, n=1, \dots, T\}$ is to first consider the process $\{Z_n, n=1, \dots, T\}$ defined in Proposition \ref{prop:granger_proy} and use the identity \eqref{eq:granger_proy}. Thus, the process described below aims to adapt the approach used in \citet{seo_functional_2024} to the multivariate case, but with greater detail and adapted to our purposes. 

First, we define an orthogonal matrix $[\beta, \beta_\perp] \in \mathbb{R}^{m \times m}$ such that $\beta \in \mathbb{R}^{m \times r}$ and $\beta_\perp \in \mathbb{R}^{m \times (m - r)}$ for a fixed value of $r \le m$.

For simplicity, we consider $Z_0 = 0$ and the following recursion:
\begin{align*}
  Z_{n+1}^N &= \left(\beta_{\perp} \text{diag}(\bm{\gamma}) \beta_{\perp}'\right) Z_n \\
  Z_{n+1}^S &= \left(\beta \text{diag}(\bm{\alpha}) \beta'\right) Z_n \\
  Z_{n+1} &= Z_{n+1}^N + Z_{n+1}^S + \epsilon_{n+1},
\end{align*}
for $n = 1, \dots, T - 1$, where $\bm{\alpha} = (\alpha_1, \dots, \alpha_{m - r})'$, $\bm{\gamma} = (\gamma_1, \dots, \gamma_r)'$, and $\abs{\alpha_j} < 1$, $\abs{\gamma_j} < 1$. The process $\epsilon$ is assumed a white noise with mean $0$ and covariance matrix $\Sigma$. Thus, $\{Z_n, n \ge 0\}$ is a stationary VAR$(1)$ process. 

Furthermore, to provide generality to the simulated scenarios, the innovation process $\epsilon$ is drawn from a Student’s $t$-distribution with 3 degrees of freedom, constructed using the \texttt{portes} package \citep{mahdi_portes_2023}, and a general covariance matrix $\Sigma$ is obtained using the function $\texttt{genPositiveDefMat}$ from the \texttt{clusterGeneration} package \citep{qiu_clustergeneration_2023}.

Therefore, for building a process under the last scenarios, we do the following  for each of them:
\begin{itemize}
  \item \textit{Scenario 1:} In this case, we use directly the decomposition described in Proposition \ref{prop:granger_proy}:
\[
P_{\beta_{\perp}} \Delta X_n = Z_n^N \quad \text{and} \quad P_{\beta} X_n = Z_n^S,
\]
so that
\[
X_n := \sum_{k=0}^{n} Z_k^N + Z_n^S.
\]

In this way, we construct $\{X_n, n \ge 0\} \sim \textsf{CI}(1,1)$.

  \item \textit{Scenario 2:} Unlike the previous scenario, here the components of $\{X_n, n=1, \dots, T\}$ will not all have the same order of integration.

  In practice, processes up to order $\textsf{I}(2)$ are typically considered. Thus, we define $M_1, M_2 \subset \{1, \dots, m\}$ such that $M_1 \cap M_2 = \emptyset$, as index sets identifying the components of $\{X_n, n=1, \dots, T\}$ that are $\textsf{I}(1)$ and $\textsf{I}(2)$, respectively.

  In this scenario, we define
  \[
    X_{i,n} = \begin{cases}
      \sum_{k=0}^{n} Z_{i,k}^N + Z_{i,n}^S & \text{if } i \in M_1, \\
      \sum_{k=0}^{n} \sum_{p=0}^{k} Z_{i,p}^N + Z_{i,n}^S & \text{if } i \in M_2, \\
      Z_{i,n} & \text{otherwise}.
    \end{cases}
  \]
\end{itemize}

Afterwards, an estimator $\hat{\beta}$ is obtained using the Johansen, PCA, and PLS procedures. In addition, we also consider the SPCA-based methodology. Specifically, we use the implementation in the package \texttt{elasticnet}\citep{zou_elasticnet_2020} which uses the approach in \cite{zou_sparse_2006}. This implementation provides the option to select the number of non-zero entries for each of the estimate components. Although the problem of adequately choosing the penalty parameter requires a separate investigation, as it relies on the application and is mainly thinking fora high-dimensional settings. Finally, to have a fair comparison, we specify that we also used a centred and scaled sample prior to applying the PCA-based methodologies -- Algorithm~\ref{algo:PLS_TS} assumes that for PLS methodology. 

On the other hand, it is necessary to evaluate the error between the estimated vector subspace in each methodology and the theoretical one. To quantify this error, we consider the generalised Grassmann distance:
\[
\delta(\hat{\beta},\beta):=\left(\sum_{i=1}^{\min\{r,\hat{r}\}}\theta_i^2+\abs{\hat{r}-r}\left(\frac{\pi}{2}\right)^2\right)^{\frac{1}{2}},
\]
where $\theta_i$ is the $i$-th principal angle; more details can be found in \citet{ye_schubert_2016}. This distance can be used to measure the discrepancy between two subspaces of possibly different dimensions, in which case it directly penalises the dimension difference. We shall refer to this measure of discrepancy between the estimated and theoretical subspace as the subspace estimation error.

The simulation study is constructed by considering a value of $m = 11$, for which all methods can be applied. We refer to this as the low-dimensional case, and we choose $m = 11$ because it is the upper limit for Johansen. For the SPCA-based methodology in the low-dimensional setting, we consider a penalised parameter of $0.25$ for each coordinate. The intention is to illustrate the effect of incorporating slight sparsity. We also consider $m = 300$, where only PLS, PCA and SPCA can be applied, and we refer to this as the high-dimensional scenario. In this case, we also set up the penalised parameter 0.25, and use the KPSS test for the PLS- and PCA-based methodologies for the reasons previously discussed.

For $m = 11$, we consider $r = 10, 9, 8$, and $T = 100$. In this way, the associated stable space has a sufficiently large dimension to ensure that the corresponding projection is informative, while also allowing us to assess the effect of dimensionality on estimation. Similarly, by adjusting the scale of the values, for $m = 300$, we consider $r = 250, 200, 150$, and $T = 100$. This allows us to evaluate the PLS- and PCA-based methodologies in a high-dimensional setting. Regarding the order of integration of the series, for each value of $m$ and $r$, we evaluate the following cases:
\begin{enumerate}
    \item For $m = 11$ and the given value of $r$:
    \begin{enumerate}
      \item \textit{Case 1)}: $\abs{M_1} = 11$, that is, $\{X_n, n=1, \cdots, T\} \sim \textsf{CI}(1,1)$.
    
      \item \textit{Case 2)}: $\abs{M_1} = 10$ and $\abs{M_2} = 0$.
    
      \item \textit{Case 3)}: $\abs{M_1} = 9$ and $\abs{M_2} = 1$.
    
      \item \textit{Case 4)}: $\abs{M_1} = 8$ and $\abs{M_2} = 2$.
    \end{enumerate}

    \item For $m = 300$ and the given value of $r$:
    \begin{enumerate}
      \item \textit{Case 1)}: $\abs{M_1} = 300$, that is, $\{X_n, n=1, \cdots, T\} \sim \textsf{CI}(1,1)$.
    
      \item \textit{Case 2)}: $\abs{M_1} = 250$ and $\abs{M_2} = 10$.
    
      \item \textit{Case 3)}: $\abs{M_1} = 200$ and $\abs{M_2} = 20$.
    
      \item \textit{Case 4)}: $\abs{M_1} = 150$ and $\abs{M_2} = 30$.
    \end{enumerate}
\end{enumerate}

To define the sets $M_1, M_2 \subset \{1, \dots, m\}$, we can do so without loss of generality in an ordered fashion. That is, for \textit{Case 3} with $m = 11$, since $\abs{M_1} = 5$, we take $M_1 = \{1, 2, \dots, 5\}$ and $M_2 = \{6, 7\}$.

For the low-dimensional case, we perform $S = 500$ simulations with $T = 100$, fixing the parameters $m, r, \beta, \Sigma, \bm{\alpha}$, and $\bm{\gamma}$. For the high-dimensional analysis, we conduct $S = 100$ simulations due to computational constraints. In each simulation, we obtain the error in estimating the dimension, reporting the average value (with standard deviation in parentheses) of these errors across the $S$ simulations.
\begin{table}[h!]
    \centering
\begin{tabular}{cccccc}
  \hline
Method & r & Case 1 & Case 2 & Case 3 & Case 4 \\ 
  \hline
Johansen & 10.00 & 3.799 (0.527) & 3.692 (0.516) & 3.493 (0.330) & 4.443 (0.233) \\ 
  PCA & 10.00 & 0.899 (0.470) & 0.995 (0.434) & 1.948 (0.744) & 1.594 (0.547) \\ 
  PLS & 10.00 & 0.862 (0.475) & 0.972 (0.460) & 1.813 (0.552) & 1.574 (0.396) \\ 
  SPCA & 10.00 & 2.978 (0.968) & 2.697 (0.973) & 3.139 (0.713) & 3.161 (0.647) \\ \hline
  Johansen & 9.00 & 3.454 (0.519) & 3.479 (0.500) & 3.033 (0.438) & 3.084 (0.426) \\ 
  PCA & 9.00 & 1.395 (0.483) & 1.428 (0.460) & 1.620 (0.468) & 1.483 (0.391) \\ 
  PLS & 9.00 & 1.177 (0.542) & 1.220 (0.553) & 1.497 (0.458) & 1.439 (0.364) \\ 
  SPCA & 9.00 & 3.384 (0.721) & 3.299 (0.693) & 3.339 (0.683) & 3.378 (0.689) \\ \hline
  Johansen & 8.00 & 3.328 (0.464) & 3.326 (0.410) & 3.023 (0.385) & 3.239 (0.438) \\ 
  PCA & 8.00 & 1.968 (0.332) & 1.987 (0.313) & 2.023 (0.280) & 2.119 (0.249) \\ 
  PLS & 8.00 & 1.826 (0.367) & 1.854 (0.349) & 1.880 (0.306) & 2.017 (0.264) \\ 
  SPCA & 8.00 & 2.716 (0.600) & 2.677 (0.584) & 2.582 (0.550) & 2.448 (0.491) \\ 
   \hline
\end{tabular}
 \caption{Comparison of Johansen procedure, PLS and PCA subspace estimation error with $m=11$ and $T=100$ across the different cases and $r$ values. We add in parentheses the standard deviation through the $S=500$ simulations.}
    \label{tab:results_low_dimension}
\end{table}

Table \ref{tab:results_low_dimension} shows the error in estimating the stable subspace for the different methods across the proposed cases with $m = 11$. We observe across the instances and values of $r$ a considerably better performance of the PCA- and PLS-based methodologies compared to the Johansen procedure.

For visualisation purposes, we also consider simulating $S = 500$ time series of length $T = 100$ with $(m, r) = (11, 9), (6, 4)$, that is, a gradual increase in dimension while preserving the same codimension 2 for the stable space. Similarly to the low-dimensional setting, we consider a penalised parameter of $0.25$ to visualise the effect of sparsity inclusion. 
\begin{figure}[!ht]
    \centering
    \includegraphics[width=0.9\textwidth]{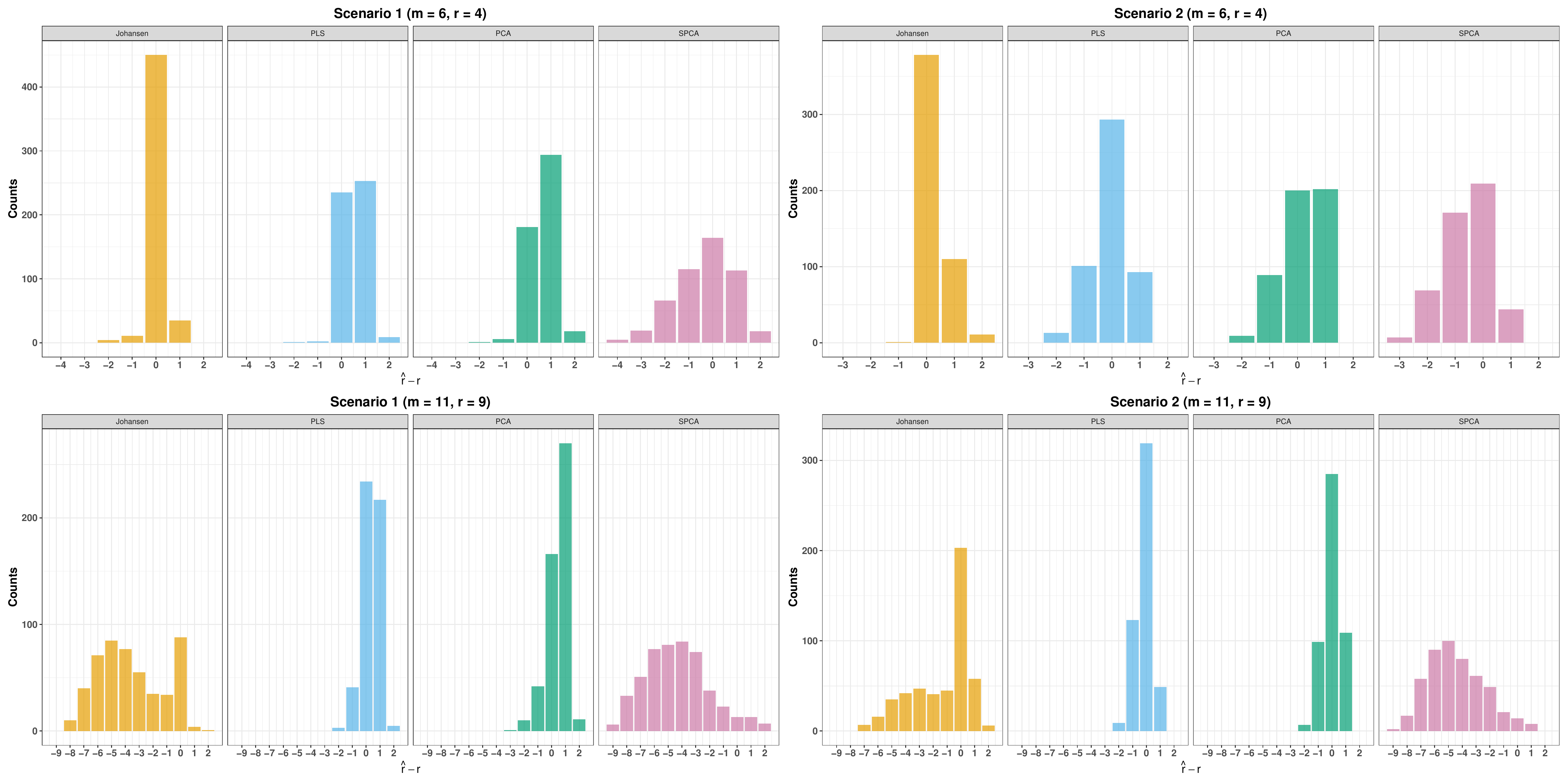}
    \caption{We contrast the differences between Scenario 1 (cointegrated series) with Scenario 2 in the dimension estimation error $\hat{r}-r$ for $(m,r)=(11,9),(6,4)$ using histograms across $S=500$ simulations.}
    \label{fig:dim}
\end{figure}

\begin{figure}[!ht]
    \centering
    \includegraphics[width=0.9\textwidth]{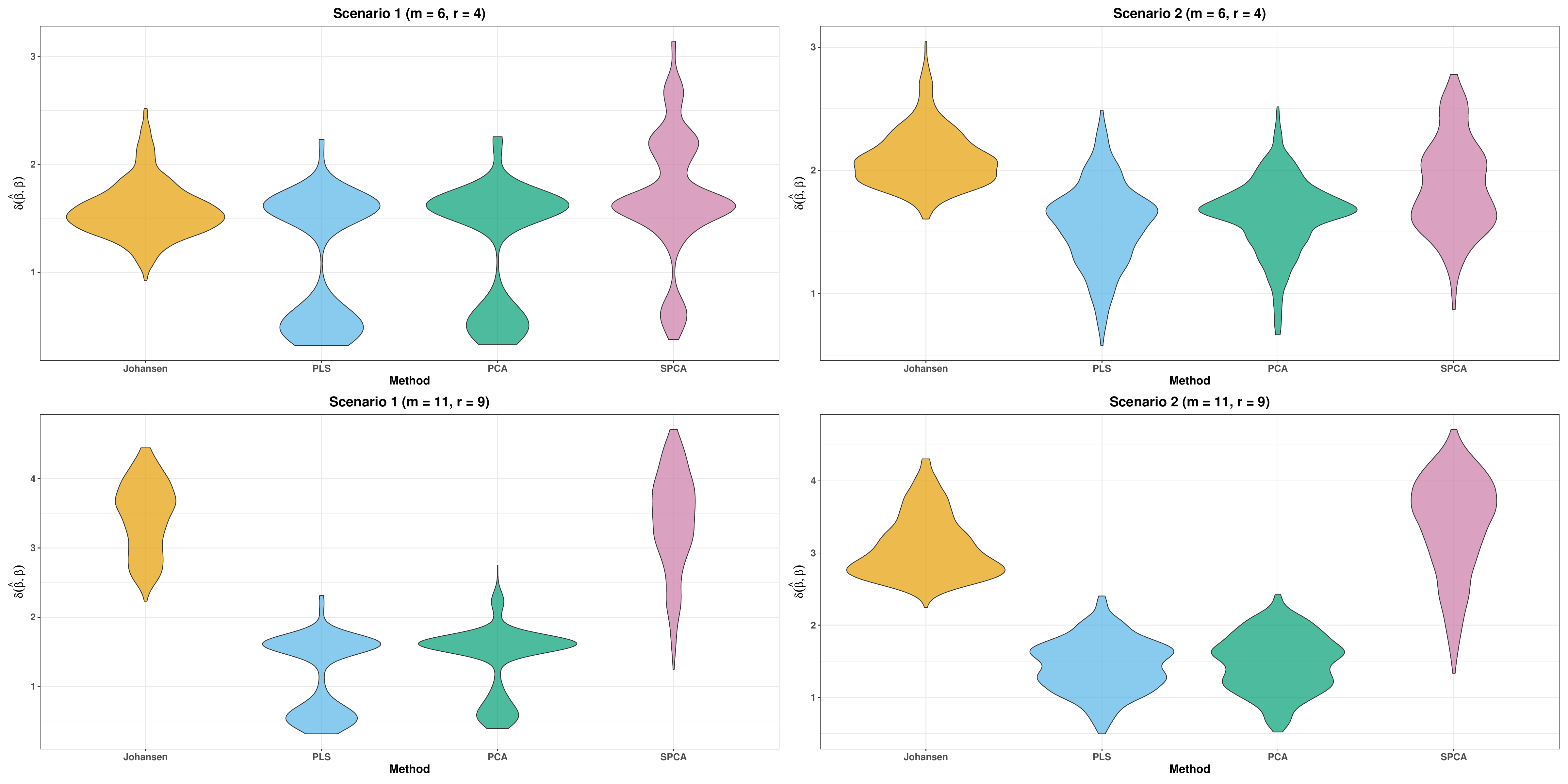}
    \caption{We contrast the differences between Scenario 1 (cointegrated series) with Scenario 2 in the stable space estimation error for $(m,r)=(11,9),(6,4)$ across $S=500$ simulations using violin diagrams.}
    \label{fig:norm}
\end{figure}

It is of interest to visualise the two scenarios previously described: the first is the cointegration scenario, and the second scenario is illustrated by considering $(\abs{M_1}, \abs{M_2}) = (m - 3, 2)$. First, in Figure \ref{fig:dim}, we visualise the distribution of $\hat{r} - r$, where positive values indicate overestimation of the dimension, and negative values indicate estimation of a stable space of smaller dimension. In Figure \ref{fig:norm}, we additionally visualise the distribution of $\delta(\hat{\beta}, \beta)$ for each methodology through violin diagrams

In both the distribution of $\hat{r} - r$ and the subspace error, under the cointegration and low-dimensional setting, we can observe more or less agreement between the methodologies, which is what we expected because it is the baseline case. However, we can see that the Johansen procedure is more sensitive to increases in dimension as well as to changes in the scenario. Regarding the SPCA-based method, we can see that incorporating a slight sparsity leads to a worse performance, although it is similar to the Johansen behaviour for $m=11$. This is because we have not simulated a sparse structure in the $\{Z_n,n=1,\dots,T\}$ process simulations. We include the SPCA method result for completeness of our analysis.  

For Case 1, which corresponds to the estimation of the cointegration space, as expected, all four methodologies yield consistent results for the different values of $r$, which can also be observed in Figure \ref{fig:dim} -- a centred histogram at 0. We observe that as the value of $m$ increases, the estimation is severely affected. This is not the case for the PLS- and PCA-based methods, which maintain a more robust estimation of $r$ as the dimension increases. Thus, we see that the Johansen procedure struggles more in determining the dimension of the stable space as the scenario changes and the dimensionality increases.

Moreover, regarding the value of $\delta(\hat{\beta}, \beta)$, Figure \ref{fig:norm} shows similar performance across the three methodologies in the cointegration context. However, the Johansen procedure again exhibits poorer performance outside this setting compared to PLS and PCA, which in turn show comparable behaviour. The deviation of SPCA from PCA method is controlled through the sparsity parameter. Here, we can notice a slight deviation in the estimation, which is similar to the Johansen performance overall. 

On the other hand, Table \ref{tab:results_high_dimension} summarises the results for $m = 300$ with the suggested values for the dimension of the stable space $r$. As in the low-dimensional case, we observe a similar difference between the PCA- and PLS-based methodologies, but in this case with slightly better performance in favour of the PCA-based method. It is interesting to observe a better performance of the SPCA-based method with the decrease of $r$, the stable space dimension, although with slightly more uncertainty.
\begin{table}[h!]
    \centering
\begin{tabular}{cccccc}
   \hline
Method & r & Case 1 & Case 2 & Case 3 & Case 4 \\ 
  \hline
PCA & 250.00 & 10.824 (0.095) & 10.657 (0.150) & 10.340 (0.225) & 10.012 (0.250) \\ 
  PLS & 250.00 & 15.884 (0.060) & 15.632 (0.064) & 15.575 (0.056) & 15.676 (0.056) \\ 
  SPCA & 250.00 & 22.532 (0.334) & 22.011 (0.244) & 21.641 (0.187) & 21.287 (0.173) \\ \hline
  PCA & 200.00 & 15.539 (0.035) & 15.466 (0.073) & 15.207 (0.170) & 14.803 (0.218) \\ 
  PLS & 200.00 & 18.128 (0.058) & 17.928 (0.039) & 17.907 (0.038) & 17.979 (0.037) \\ 
  SPCA & 200.00 & 20.479 (0.226) & 19.721 (0.205) & 19.339 (0.170) & 18.984 (0.153) \\ \hline
  PCA & 150.00 & 19.102 (0.032) & 19.051 (0.063) & 18.768 (0.152) & 18.289 (0.244) \\ 
  PLS & 150.00 & 20.348 (0.029) & 20.199 (0.035) & 20.198 (0.033) & 20.244 (0.036) \\ 
  SPCA & 150.00 & 17.597 (0.208) & 16.728 (0.185) & 16.356 (0.171) & 16.033 (0.165) \\ 
   \hline
\end{tabular}
 \caption{Comparison of PLS, and PCA subspace estimation error with $m=300$ and $T=100$ across different cases and $r$ values. We add in parentheses the standard deviation through the $S=100$ simulations.}
\label{tab:results_high_dimension}
\end{table}

Therefore, these methodologies offer a practical and efficient alternative for estimating the stable space in high-dimensional settings. To generate all tables and plots from the simulation study, we use the code provided in the Supplementary Material. In what follows, we illustrate the estimation and practical use of the stable space through real-world examples.

\section{Practical Examples}
\label{sec:examples}

\subsection{Low-dimensional example}
In the low-dimensional practical example, we analyse the behaviour of several economic indicators associated with inflation in Mexico. These variables form a vector time series $X$, and are described in Table \ref{tab:variables_inflation}.
\begin{table}[ht]
\centering
\begin{tabular}{|c|| c|}
  \hline
\textbf{Variables} & \textbf{Description} \\ 
\hline
 P & National Consumer Price Index (INEGI)\\ 
 U & Unemployment Rate (INEGI)\\ 
 BYM & Banknotes and Coins (Banxico)\\ 
 E & Nominal Exchange Rate (INEGI)\\ 
 R & 28-Day Interbank Interest Rate (Banxico)\\ 
 W & Real Wages (Banxico)\\ 
 PUSA & United States National Consumer Price Index (BLS)\\ 
   \hline
\end{tabular}
\caption{Variables that make up the vector series related to inflation in Mexico. In parentheses, we indicate the public source from which the indicator comes.}
\label{tab:variables_inflation}
\end{table}

The data, obtained from the National Institute of Statistics and Geography (INEGI, by its Spanish acronym) in Mexico, Banxico (Mexico’s central bank), and the U.S. Bureau of Labour Statistics (BLS), cover the period from January 2005 to January 2023, with monthly frequency. The compiled dataset is available in our GitHub repository; see \ref{append:software}. This yields a random sample $\{X_n, n = 1, \dots, T\}$, with $m = 7$ and $T = 216$. 

Following the information in Table~\ref{tab:variables_inflation}, where  P denotes the price level, each time series can be interpreted in relation to P as follows: U is associated with a Phillips curve relationship; BYM and R capture the link between prices and excess money in circulation, with the interest rate acting as an anchor to contain inflation. E and W represent input and wage costs faced by firms, while PUSA may be related to imported inflation. A similar study in this context is \cite{bailliu_explaining_2003}, although it does not consider all these indicators jointly.

It is important to emphasise that no structural relationships are imposed in this analysis; instead, we explore the dynamics through both stable and unstable combinations across the panel of time series. To approximate Assumption~\ref{ass:nondeterministic} as closely as possible, a preliminary analysis is conducted in which both the seasonal component and a linear trend are removed, as outlined in that assumption. For the BYM variable, the analysis is performed on the logarithmic scale.

For each of the series, a KPSS test can be conducted at the level of the series and after differencing, until there is no evidence to reject the null hypothesis. This allows for an empirical estimation of the order of integration of each series. The results are summarised in Table \ref{tab:pvalues}, which shows the p-values obtained from the KPSS test for each corresponding series.
\begin{table}[ht]
\centering
\begin{tabular}{c|ccc}
  \hline
  Variable & X & $\Delta X$ & $\Delta^2 X$ \\ 
   \hline
 P & 0.01 & 0.04 &  \checkmark \\ 
  U & 0.02 & \checkmark & - \\ 
  BYM & \checkmark & - & - \\ 
  E & \checkmark & - & - \\ 
  R & 0.02 & \checkmark & - \\ 
  W & 0.01 & 0.04 & \checkmark \\ 
  PUSA & \checkmark & - & - \\ 
   \hline
\end{tabular}
\caption{p-value of the KPSS test applying to each coordinate of the original series $\{X_n,n=1,\dots,T\}$ and considering one and two differences. The \checkmark indicates when the null hypothesis was not rejected using a $5\%$ significance level, and the - indicates that a hypothesis test was not done.}
\label{tab:pvalues}
\end{table}
From Table \ref{tab:pvalues}, we can deduce the following:
\begin{itemize}
  \item The variables BYM, E, and PUSA correspond to $\textsf{I}(0)$ series.
  \item The series associated with variables U and R are $\textsf{I}(1)$.
  \item The variables P and W are $\textsf{I}(2)$ series.
\end{itemize}
\begin{figure}[!ht]
  \centering
  \includegraphics[width=0.8\textwidth]{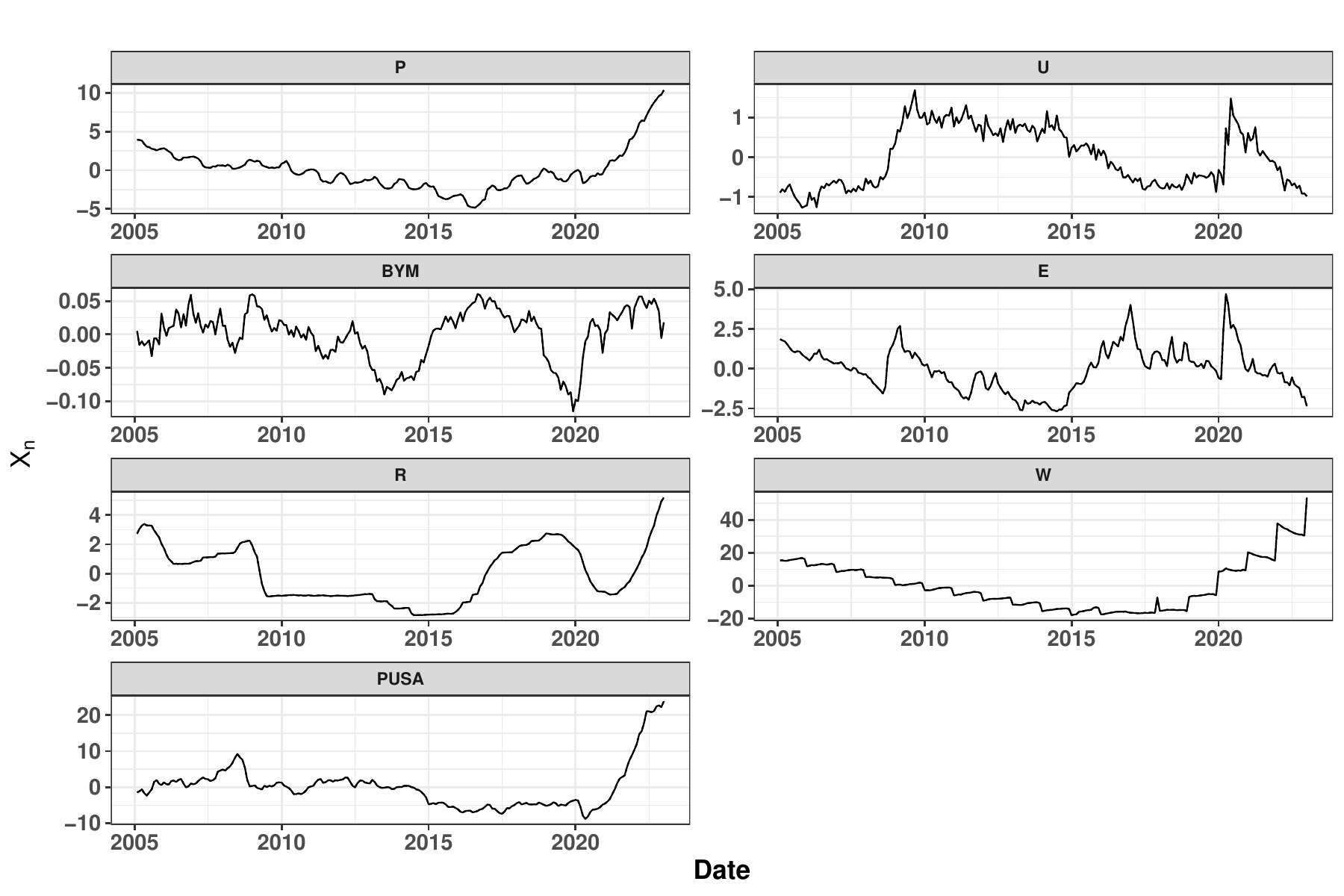}
  \caption{Visualisation of each time series corresponding to the economic indicators in Table \ref{tab:variables_inflation}.}
  \label{fig:series_inflation-png}
\end{figure}

Thus, there is evidence indicating that the vector series formed by these variables has components with different orders of integration, i.e., this example falls under Scenario 2. In Figure \ref{fig:series_inflation-png}, we can observe the dynamics of each of these economic indicators after preprocessing.

We estimate a basis for the stable space using each of the three methodologies for the economic indicator series $\{X_n, n=1, \dots, T\}$. It is important to note that the stationary scores also decompose the information into orthogonal factors; therefore, it is of interest to understand the contribution of each variable in constructing these scores. 

\begin{figure}[!ht]
  \centering
  \includegraphics[width=\textwidth]{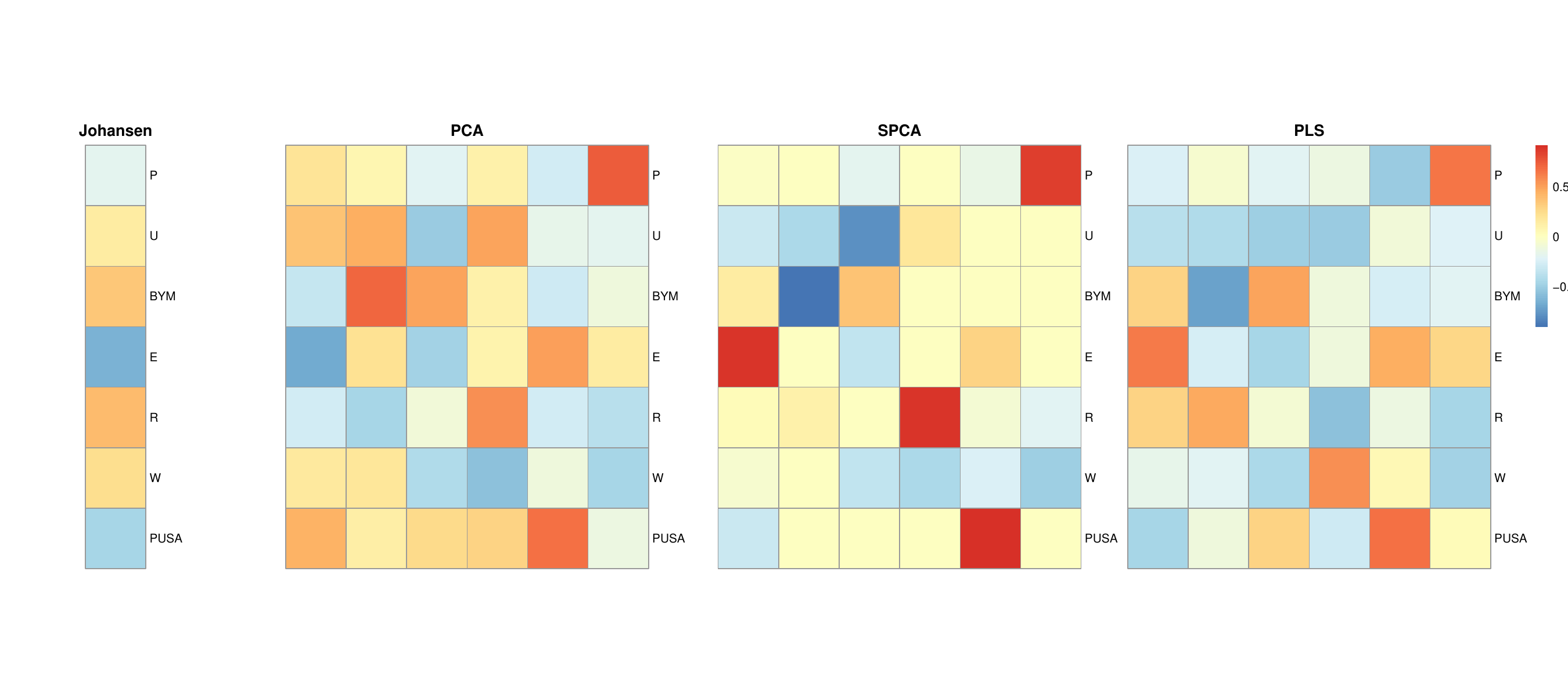}
  \caption{Heatmap that represents the weights of each variable in Table \ref{tab:variables_inflation} on the stable scores construction.}
  \label{fig:weights_inflation}
\end{figure}

For the Johansen, PCA and SPCA procedures -- where a penalised parameter of 1 is considered --, the columns of $\hat{\beta}$ obtained are directly the weights that define the corresponding scores at the series level. However, for the PLS-based methodology, the columns of $\hat{\beta}$ must be transformed as described in Proposition \ref{prop:pesosBeta}. Taking this into account, we can visualise the weights associated with each economic indicator by component in Figure \ref{fig:weights_inflation}.

We observe that for the first component in each methodology, the highest weight is on variable E. We can see that clearly in the SPCA methodology, where the sparsity helps to highlight better the variables considered for each component. For the PLS- and PCA-based methods, the estimated dimension of the stable space is 6, whereas the Johansen procedure only identifies one stable relation. This exemplifies the effect of misspecifying cointegration over a panel of time series. Although we have validated PLS and PCA in the simulation study, which yield similar results, we observe that the components assign different weights to the variables in constructing the corresponding scores. This may affect interpretation, particularly for some components. We observe that the key difference between PLS and PCA lies in the sign. This could not be seen in the simulation study, as the metrics used to evaluate the methodologies are sign-invariant.

In general, the problem of estimating the associated signs is challenging. For instance, \citet{horvath_inference_2012} notes that in PCA, signs cannot be calculated from the data. This opens up the possibility of formalising whether this is feasible with PLS. Indeed, as shown, for example, in \citet{bastien_pls_2005}, the coordinates $\hat{w}_{i}$ correspond to the partial correlations between $X_n$ and $X_{n-1}$ given $T_1, \dots, T_{i-1}$, which implies that the weights obtained via PLS have a direct interpretation in terms of dependence between $X_n$ and $X_{n-1}$. This is not the case for the PCA-based methodology, even if both methods estimate the same space. Additionally, we can see that indeed the SPCA modification offers an attractive advantage for interpretation. In Figure \ref{fig:weights_inflation}, we can see that the sparsity inclusion helps to identify which variables are more important for a specific score.

On the other hand, PCA and PLS are dimensionality reduction methods. In practice, the first components are selected based on the idea that they capture the greatest variability of the observed information and for visualisation purposes. This idea can also be applied here, with the additional consideration that these components should yield stationary scores, which is conceptually more appropriate.

Thus, we compare the projection error of the PCA- and PLS-based methodologies with the Johansen procedure in a dimensionality reduction context. For each method, we select the first two stationary scores, since only two components are available for the Johansen procedure. We then compute the projection of the original series onto the stationary scores using a regression model. In the PLS model, this regression is equivalent to the decomposition in \eqref{eq:PLS_decomposition}.

Additionally, we consider the outcome of selecting the first two components without assessing whether they yield stationary scores. We illustrate this using PCA and refer to this method as classical PCA, in contrast to the proposed approach, which involves selecting the leading stable components and is referred to as stationary PCA (or stationary PLS, where applicable). In this context, the classical PCA approach corresponds to the selection of one non-stationary component associated with the first principal component, along with the first stationary component.

After selecting the corresponding scores and performing the projection, we compute the error between the observed series and its projection. Specifically, we calculate the mean squared error (MSE) for each series, normalised with respect to its sample variance.

\begin{table}[ht]
\centering
\begin{tabular}{c|c|c|c|c|c}
 \hline
Variable & Johansen & classic PCA & stationary PCA & stationary SPCA & stationary PLS \\ 
  \hline
P & 0.91 & 0.10 & 0.26 & 0.32 & 0.26 \\ 
  U & 0.92 & 0.62 & 0.98 & 0.18 & 0.99 \\ 
  BYM & 1.00 & 0.86 & 0.79 & 0.98 & 0.80 \\ 
  E & 0.85 & 0.27 & 0.44 & 0.46 & 0.45 \\ 
  R & 0.87 & 0.41 & 0.95 & 0.14 & 0.95 \\ 
  W & 0.69 & 0.07 & 0.17 & 0.49 & 0.16 \\ 
  PUSA & 0.98 & 0.09 & 0.06 & 0.09 & 0.06 \\ 
   \hline
\end{tabular}
\caption{Normalised MSE of the projection onto the main two stationary scores for each method, and against considering the first two principal components regardless of their stability}
\label{tab:projection_error_low_dimension}
\end{table}
\begin{figure}[!ht]
    \centering
    \includegraphics[width=0.8\textwidth]{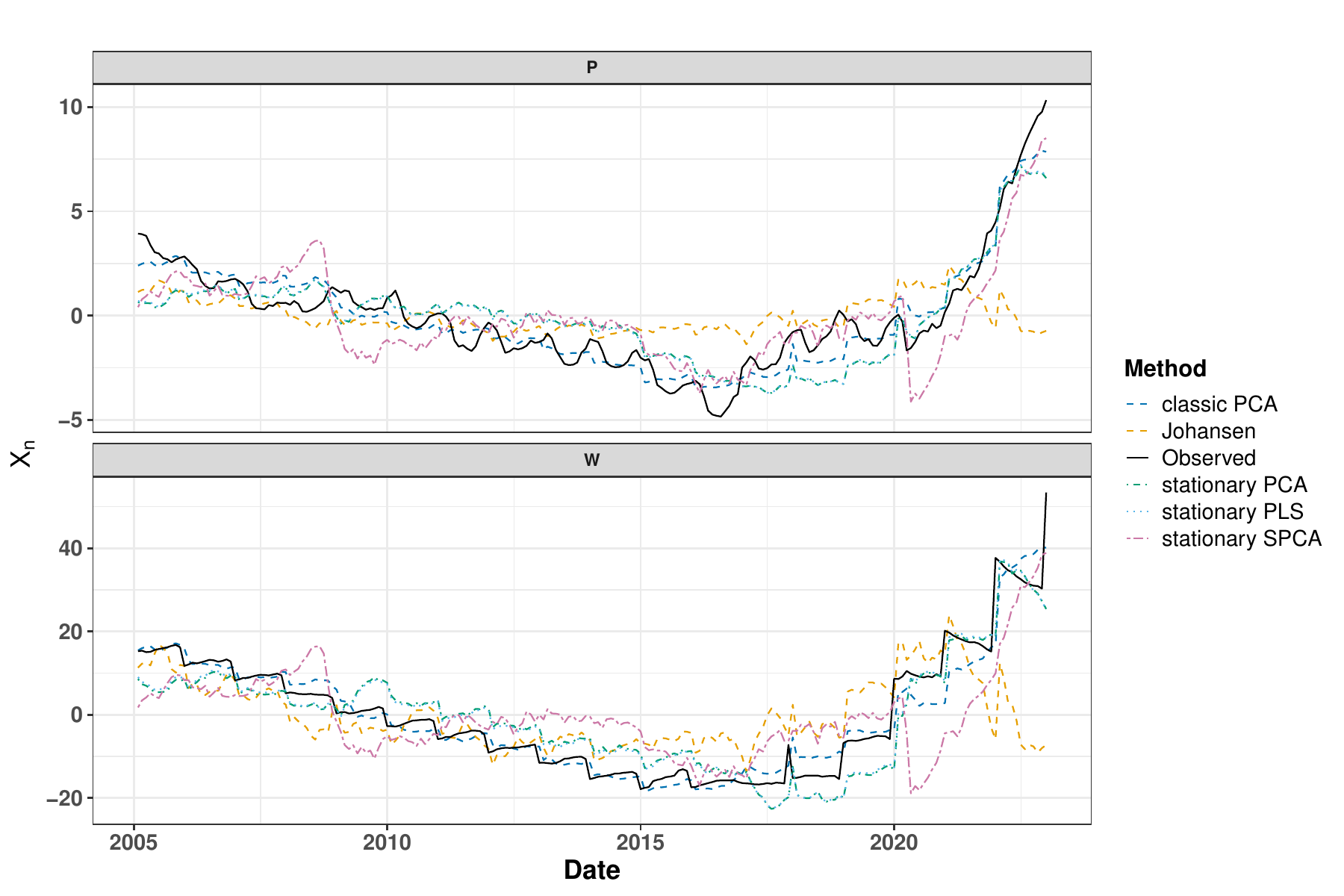}
    \caption{Visualisation of the scaled P and W indicators projection onto the two main stability components. For PCA, we also consider the projection on the two main principal components regardless of stability.}
    \label{fig:projection_low_dimension}
\end{figure}
These errors are reported in Table \ref{tab:projection_error_low_dimension}. We observe that the Johansen procedure performs worse compared to the other methodologies across all series. On the other hand, PLS and stationary PCA yield similar results, while classical PCA -- although showing better performance -- does so by capturing the non-stationary aspects of the observed sample, due to the non-stability of the first principal component.

Finally, in Figure \ref{fig:projection_low_dimension}, we visualise the projection for each of the methodologies and approaches, for example, focusing on the variables P and W, where the difference in error between the Johansen procedure and PCA/PLS is considerable. Moreover, we can observe the non-stationary aspects captured by including the first non-stable principal component in classical PCA, and the contrast with selecting the first two stable components.

Although classical PCA provides a good fit within the sample range, stationary PCA exhibits extrapolation capability and captures the most relevant stationary part of each series, yielding a better fit in some regions than in others. For the SPCA modification, we can observe that increasing the penalty results in a more spiky behaviour with respect to the baseline PCA. If the objective is estimation and forecasting, the penalty parameter should be controlled cautiously, taking care of the trade-off between fitting and variable selection. Moreover, the areas where the difference between the stationary projection and the observed series is largest could be interpreted as periods where the process deviates most from stationarity, opening the door to investigating the underlying causes.

\subsection{High-dimensional example}

We now analyse $m = 205$ monthly series retrieved from the indicators database provided by INEGI. Specific information about these series can be found in the catalogues available in the repository referenced in \ref{append:software}, along with the dataset itself. Some of the variables included in this database are the unemployment rate, indicators of economic activity in the construction sector (variable \texttt{s23} in the dataset), and the professional, scientific, and technical services sector (variable \texttt{s54} in the dataset). 

We apply the same preprocessing scheme to the data to obtain a process that satisfies, at a minimum, Assumption~\ref{ass:nondeterministic}. The information for each of the 205 economic indicators covers the period from January 2017 to December 2024, that is, $T = 95$. Therefore, we are in the case where $T < m$. In the same way as in the low-dimensional setting, we identify the order of integration of each series using the KPSS test. Among the 205 series in the system, we find that 185 are identified as $\textsf{I}(0)$, 19 as $\textsf{I}(1)$, and one as $\textsf{I}(2)$.

Thus, we would expect the estimated dimension of the stable space to be at least 185. However, it could be higher if there are components that compensate across all series -- including the non-stationary ones-- to produce stationary scores. The estimators $\hat{\beta}_{\text{PLS}},$ $\hat{\beta}_{\text{PCA}}$ and $\hat{\beta}_{\text{SPCA}}$ are computed using a significance level of $5\%$ for detecting the stable components and considering a penalised parameter of 1 for the SPCA modification. We obtain $\dim \text{Col }\hat{\beta}_{\text{PLS}} = 191$, while $\dim \text{Col }\hat{\beta}_{\text{PCA}} = 196$ and $\dim \text{Col }\hat{\beta}_{\text{SPCA}} = 81$, so the dimensions of the estimated non-stable spaces are 14, 9, and 124, respectively.

In this case, for both the PLS, PCA and SPCA methodologies, the first two stable components coincide with the first two components obtained by each method, i.e., the stationary approach aligns with the classical one. This illustrates that when Assumption \ref{ass:cointegration} does not hold, it is necessary to identify the stable components explicitly, even if they do not appear in order. Therefore, we see that although the multivariate series includes non-stationary components, the majority of series identified as $\textsf{I}(0)$ dominate the variance/covariance structure of the system. As a result, the first components obtained by the methodologies are stable.

To illustrate the difference between stable and non-stable components, we consider the approximation of the observed data using the first two components in each case. When using the stable components, we refer to the stationary approach. In contrast, the approximation based on the first two non-stable components is referred to as the non-stable version of the method. We observe that the stationary approach yields a slightly greater error than the non-stationary one, even though all the series shown in Table \ref{tab:projection_error_high_dimension} are non-stationary. Consequently, the first two stable components -- which in this case coincide with the first two overall components -- not only load on the stationary components but also on the non-stationary ones in such a way as to construct the corresponding stationary scores. This last phenomenon somewhat contradicts the previous example. It may be due either to the larger number of $\textsf{I}(0)$ components, as previously mentioned, or to the curse of dimensionality. We observe the same phenomena incorporating regularisation techniques in the form of the SPCA method.

\begin{table}[ht]
\centering
\begin{tabularx}{\textwidth}{c|X|X|X|X|X|X}
  \hline
 Variable & non-stable PCA & stationary PCA & non-stable PLS & stationary PLS & non-stable SPCA & stationary SPCA \\ 
  \hline
s23 & 0.30 & 0.43 & 0.36 & 0.44 & 0.48 & 0.33 \\ 
imaief\_Hgo & 0.43 & 0.64 & 0.42 & 0.60 & 0.60 & 0.60 \\ 
imaief\_SLP & 0.52 & 0.54 & 0.53 & 0.54 & 0.57 & 0.53 \\ 
s54 & 0.52 & 0.58 & 0.63 & 0.61 & 0.68 & 0.43 \\ 
rem\_cmenor & 0.30 & 0.44 & 0.45 & 0.47 & 0.55 & 0.20 \\ 
  \hline
\end{tabularx}
\caption{Normalised MSE of the projection onto the main two stationary scores and the two main non-stable components on some series in the system. The \textsf{rem\_cmenor} is the $\textsf{I}(2)$ series whereas the rest are $\textsf{I}(1)$ series.}
\label{tab:projection_error_high_dimension}
\end{table}

In Table \ref{tab:projection_error_high_dimension}, we present the normalised MSE for several series in the system. \textsf{s23} represents the construction indicator, \textsf{s54} corresponds to professional and scientific services, \textsf{imaief\_Hgo} and \textsf{imaief\_SLP} represent industrial activity indicators for the Mexican states of Hidalgo and San Luis Potosí, respectively, and \textsf{rem\_cmenor} is the retail trade indicator.
\begin{figure}[!ht]
    \centering
    \includegraphics[width=0.8\textwidth]{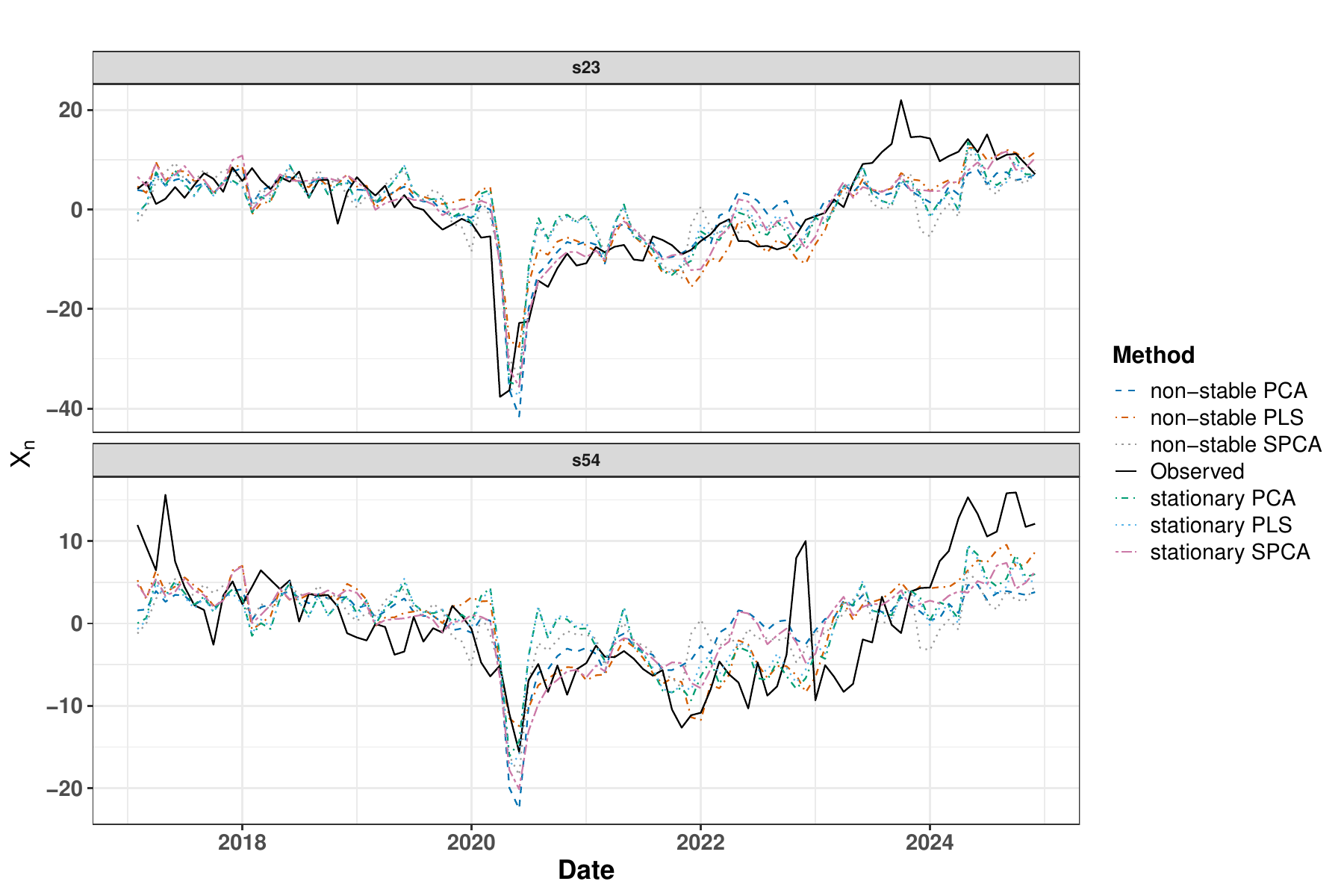}
    \caption{Visualisation of the scaled \textsf{s23} and \textsf{s54} indicators projection onto the two main stability components and the first two non-stable components.}
    \label{fig:projection_high_dimension}
\end{figure}

As in the previous section, in Figure \ref{fig:projection_high_dimension} we visualise the projection for the series \textsf{s23} and \textsf{s54}, confirming that the adjustment provided by the stable components -- which combine all the information -- offers an acceptable approximation to the observed series even using non-stable approaches are slightly better due to the non-stationarity of the considered series. It is interesting to notice that stationary SPCA offers a better fit for the series in Table \ref{tab:results_high_dimension} than the non-stable analogue. The data and the code for reproducing the results can be found in the Supplementary Material.

\section{Discussion}
\label{sec:discussion}
The cointegration framework proposed by \citet{engle_co-integration_1987} and formalised in the Johansen procedure \citep{johansen_estimation_1991} constitutes one of the main tools for addressing non-stationarity in multivariate time series. An extension of these ideas in higher order systems is treated in \cite{granger_investigation_1989, engsted_multicointegration_1999} with the concept of multicointegration. However, as demonstrated, its practical applicability in high-dimensional settings is limited due to its computational complexity, its parametric nature, and -- as shown in this work -- the restrictive assumption of a cointegrated or multicointegrated system of series.

In contrast, methodologies based on non-parametric dimensionality reduction techniques, such as PCA and PLS -- and the SPCA proposed here, especially for high-dimensional settings -- offer a more flexible and efficient way to address the inherent complexity of systems with a large number of variables as well as interpretability of the components/scores estimated. While these methodologies can initially be applied within a cointegration framework, we have shown that they are capable of recovering the most relevant aspects of the underlying stationary dynamics through stationary scores in a more flexible setting.

The concept of a stable space, introduced in this work, is proposed as a distinct concept from the notions of cointegration and multicointegration, with a focus on practical applications. This notion allows the identification of linear combinations of the original series that generate processes with stationary behaviour. In doing so, it broadens the range of situations in which it is possible to apply analytical techniques based on the assumption of stationarity, enhancing both the applicability and robustness of statistical procedures.

A key methodological proposal of this work is the use of components associated with the stable space in the context of dimensionality reduction. While these components do not necessarily correspond to those that maximise explained variance -- as in classical principal components analysis  -- they offer the significant advantage of being stationary. This property provides, on the one hand, an alternative and lower-complexity representation of the data, and on the other, the ability to apply statistical techniques that require stationarity, such as VAR models, forecasting methods, or changepoint detection techniques.

The results of the simulation study, carefully designed to assess the performance of the different methodologies under a variety of scenarios (including explicit cointegration structures, low- and high-dimensional configurations, and partially violated assumptions), support the effectiveness of the proposed approach. In particular, the estimation of the stable space via PCA and PLS proves robust across the proposed scenarios, which represent deviations from classical assumptions. In addition, we consider regularisation techniques in the form of the SPCA for improving interpretability, and it might be used for variable selection in practical settings. Furthermore, these methods are capable of capturing relevant information in a parsimonious manner.

Finally, it is important to highlight that the approach developed in this work opens multiple avenues for future research. One of the most promising consists in generalising the concept of stable space to nonlinear contexts or more complex forms of temporal dependence. Moreover, it is crucial to address the sensitive aspect of misidentifying stationary components, specifically by controlling the false positive rate of stationary scores. This is important for quantifying the uncertainty of the stationary assumption in the score that might be used as an input for another method.  In this regard, the core idea behind the PLS algorithm -- based on capturing linear dependence between $X_n$ and $X_{n-1}$ -- provides a natural starting point for building nonlinear extensions. It is also of great interest to explore the application of this approach in specific domains in conjunction with other methodologies -- such as forecasting or changepoint detection -- in areas such as macroeconomics, financial analysis, or the study of multivariate biological systems, where the simultaneous presence of high dimensionality and non-stationarity is common.

%


\bibliographystyle{elsarticle-harv}
\bibliography{references}

\clearpage
\appendix
\setcounter{figure}{0}
\section{Proofs of Theoretical Results}
\label{appen:proofs}

\textbf{Proof of Proposition \ref{prop:granger_proy}}: This proof is based on that given in \citet{lutkepohl_new_2005}, Proposition 6.1, and follows the same notation.

First, we consider the parametrisation of the polynomial $\theta$ as
\[
\theta(z) = \theta(1) + (1 - z)\theta^{\star}(z),
\]
so that
\begin{equation}
\label{eq:reparametrisation_Z}
Z_n = \theta(1)\epsilon_n + \theta^{\star}(B)\Delta \epsilon_n.
\end{equation}

On the other hand, since $P_{\beta_{\perp}}Z_n = P_{\beta_{\perp}}\Delta X_n$, we have
\begin{align*}
X_n &= \sum_{i=1}^n \Delta X_i \\
    &= \sum_{i=1}^n (P_{\beta}\Delta X_i + P_{\beta_{\perp}}\Delta X_i) \\
    &= \sum_{i=1}^n P_{\beta_{\perp}}Z_i + \sum_{i=1}^n P_{\beta}\Delta X_i \\
    &= \sum_{i=1}^n P_{\beta_{\perp}}Z_i + P_{\beta}X_n.
\end{align*}

From the last expression and the reparametrisation \eqref{eq:reparametrisation_Z}, we obtain:
\begin{align}
X_n &= P_{\beta_{\perp}}\theta(1)\sum_{i=1}^n \epsilon_i + P_{\beta_{\perp}}\theta^{\star}(B)\sum_{i=1}^n \Delta \epsilon_i + P_{\beta}X_n \nonumber \\
    &= P_{\beta_{\perp}}\left(\theta(1)\sum_{i=1}^n \epsilon_i\right) + P_{\beta_{\perp}}\theta^{\star}(B)\epsilon_n + P_{\beta}X_n \label{eq:decomposition_cointegration}.
\end{align}

Let $\eta_n := P_{\beta_{\perp}}\theta^{\star}(B)\epsilon_n$. Note that $\{\eta_n, n \in \mathbb{N}\}$ is a stationary time series, and from this and \eqref{eq:decomposition_cointegration}, the result follows. \qed

\vspace{1cm}
\textbf{Proof of Lemma \ref{lem:simplified_optimisation}}: First, note that
\begin{equation}
\label{eq:first_optimization}
(\hat{w}_1, \hat{c}_1) = \argmax_{d \in \mathbb{R}^m} \argmax_{e \in \mathbb{R}^m} \cov(\mathbf{X}d, \mathbf{Y}e)^2.
\end{equation}

Now, let us fix $d$ with $\lVert d \rVert = 1$ and consider the optimisation problem
\[
\hat{c}(d) = \argmax_{e \in \mathbb{R}^m} \cov(\mathbf{X}d, \mathbf{Y}e)^2 \quad \text{subject to } \lVert e \rVert = 1,
\]
that is, we fix one variable and maximise with respect to the other.

On the other hand, observe that
\begin{align*}
\cov(\mathbf{X}d, \mathbf{Y}e)^2 &\propto (d' \mathbf{X}' \mathbf{Y} e)^2 \\
                                 &= \langle \mathbf{Y}' \mathbf{X}d, e \rangle^2,
\end{align*}
where $\langle \cdot, \cdot \rangle$ denotes the inner product in $\mathbb{R}^m$.

By the Cauchy–Schwarz inequality:
\[
\langle \mathbf{Y}' \mathbf{X}d, e \rangle^2 \leq \lVert \mathbf{Y}' \mathbf{X}d \rVert^2 \cdot \lVert e \rVert^2 = \lVert \mathbf{Y}' \mathbf{X}d \rVert^2,
\]
since $\lVert e \rVert = 1$.

As $d$ is fixed and the right-hand side of the inequality does not depend on $e$, the maximum of $\cov(\mathbf{X}d, \mathbf{Y}e)^2$ is attained when $e \propto \mathbf{Y}' \mathbf{X}d$, according to the equality condition of the Cauchy–Schwarz inequality.

Therefore, the optimisation problem in \eqref{eq:first_optimization} simplifies to
\begin{align*}
\hat{w}_1 &= \argmax_{d \in \mathbb{R}^m} \cov(\mathbf{X}d, \mathbf{Y}\mathbf{Y}' \mathbf{X}d)^2 \\
          &= \argmax_{d \in \mathbb{R}^m} (d' \mathbf{X}' \mathbf{Y} \mathbf{Y}' \mathbf{X}d)^2 \quad \text{subject to } \lVert d \rVert = 1,
\end{align*}
so that $\hat{w}_1$ is the eigenvector associated with the largest eigenvalue $\lambda^2 > 0$ of the matrix $\mathbf{X}' \mathbf{Y} \mathbf{Y}' \mathbf{X}$, and
\[
\hat{c}_1 \propto \mathbf{Y}' \mathbf{X} \hat{w}_1.
\]

Moreover, if $\hat{c}_1 = k \cdot \mathbf{Y}' \mathbf{X} \hat{w}_1$ and $\lVert \hat{c}_1 \rVert = 1$, then:
\begin{align*}
1 &= k^2 \cdot \hat{w}_1' \mathbf{X}' \mathbf{Y} \mathbf{Y}' \mathbf{X} \hat{w}_1 \\
  &= k^2 \cdot \hat{w}_1' (\lambda^2 \hat{w}_1) \\
  &= k^2 \cdot \lambda^2,
\end{align*}
since $\lVert \hat{w}_1 \rVert = 1$. Thus, the proportionality constant in the relation $\hat{c}_1 \propto \mathbf{Y}' \mathbf{X} \hat{w}_1$ is $k = \lambda^{-1}$, i.e., the reciprocal of the largest singular value of the sample cross-covariance matrix $\mathbf{Y}' \mathbf{X}$. \qed

\vspace{1cm}
\textbf{Proof of Proposition \ref{prop:pesosBeta}}: This result is proven by mathematical induction on $i$, the iteration number.

For $i=1$, if $A_1 = I_m$, the identity matrix in $\mathbb{R}^m$, then $\mathbf{X}^{(1)} = \mathbf{X}^{(1)}A_1$, so the result holds in this case.

Suppose the result holds for $i$. Consider $i < m$, and we will prove the result for $i+1$.

Note that,
\[
\mathbf{X}^{(i+1)} = \mathbf{X}^{(i)}P_{i}.
\]

By the induction hypothesis, we have $\mathbf{X}^{(i)} = \mathbf{X}^{(1)}A_i$, so
\[
\mathbf{X}^{(i+1)} = \mathbf{X}^{(1)}A_iP_{i},
\]
thus, define $A_{i+1} = A_iP_{i}$ and the result holds for $i+1$, which concludes the induction.

Since $A_{i+1} = A_iP_{i}$ and $A_1 = I_m$, it follows by recursion that
\[
A_{i+1} = P_{1} \cdots P_{i} \quad \text{for } i=1,2,\dots,m-1,
\]
which gives the explicit form of the transformation used to obtain the weights in the scale of the original information $\mathbf{X}^{(1)}$.

Thus,
\begin{align*}
    T_i = \mathbf{X}^{(i)}\hat{w}_i = \mathbf{X}^{(1)}A_i\hat{w}_i = \mathbf{X}^{(1)}\hat{\gamma}_i,
\end{align*}
which is the expression of the scores as a linear combination of the original variables. \qed

\vspace{1cm}
\textbf{Proof of Remark \ref{rem:simplified_update}:} We know that for $i=0,1,\dots,m-1$
\begin{equation}
\label{eq:PLS_opt_1}
\hat{w}_{i+1} = \argmax_{d} d' \tilde{S}_{i+1}' \tilde{S}_{i+1} d \quad \text{ with } \lVert d \rVert = 1,
\end{equation}
where
\[
\tilde{S}_{i+1} := T^{-1} (\mathbf{X}^{(i+1)})' (\mathbf{Y}^{(i+1)}).
\]

A simplifying assumption of Algorithm \ref{algo:PLS_TS}, as noted in \citet{hoskuldsson_pls_1988}, is to omit the update for the response variable in step 4.

This is due to the fact that
\[
(\mathbf{Y}^{(i+1)})' T_{i+1} = \mathbf{Y}' T_{i+1},
\]
which implies
\[
\mathbf{Y}^{(i+1)} = \underbrace{\left[I_{T-1} - \sum_{j=1}^i \frac{T_j T_j'}{T_j' T_j} \right]}_{B_{i+1}} \mathbf{Y}.
\]

Furthermore, one can verify that a similar update holds for the covariates, i.e., $\mathbf{X}^{(i+1)} = B_{i+1} \mathbf{X}$, thus
\begin{align*}
    \tilde{S}_{i+1}' \tilde{S}_{i+1} &\propto (\mathbf{X}^{(i+1)})' \mathbf{Y}^{(i+1)} (\mathbf{Y}^{(i+1)})' \mathbf{X}^{(i+1)} \\
    &= \mathbf{X}' B_{i+1}' \mathbf{Y} \mathbf{Y}' B_{i+1} \mathbf{X} \\
    &= (\mathbf{X}^{(i+1)})' \mathbf{Y} \mathbf{Y}' \mathbf{X}^{(i+1)}.
\end{align*}

In this way, the algorithm is simplified by considering only the residual update for the covariate matrix $\mathbf{X}$. \qed

\vspace{1cm}
\textbf{Proof of Proposition \ref{prop:consistenciaPLS}:} The idea of the proof is to use Lemma 1 from \citet{harris_principal_1997}, that is, to relate the decomposition obtained via PCA with that obtained via the PLS algorithm.

Let \(\mathbf{X}\) and \(\mathbf{Y}\) be as in \eqref{eq:natural-choice}, and consider the estimators of the asymptotic covariances:
\[
S_{XY} := T^{-1}\mathbf{X}' \mathbf{Y} = T^{-1} \sum_{i=2}^T X_{i-1}X_i',
\]
and
\[
S_{YX} := T^{-1}\mathbf{Y}' \mathbf{X} = T^{-1} \sum_{i=2}^T X_i X_{i-1}'.
\]

By hypothesis, we know that \(\{X_n,n=1,\dots,T\}\) satisfies the conditions of the Granger Representation Theorem, so that
\[
X_n = N_n + \eta_n + X_0^{\star},
\]
with \(N \sim \textsf{I}(1)\), \(\eta \sim \textsf{I}(0)\), and \(X_0^\star\) a constant.

Since we are interested in an asymptotic result, we may assume without loss of generality that \(X_0^\star = 0\).

From the above, we obtain
\begin{equation}
\label{eq:covariance_descomp}
S_{XY} = T^{-1} \left( \sum_{i=2}^{T} N_{i-1} N_i' + \sum_{i=2}^T N_{i-1} \eta_i + \sum_{i=2}^T \eta_{i-1} N_i' + \sum_{i=2}^T \eta_{i-1} \eta_i' \right).
\end{equation}

Specifically, by Theorem \ref{thrm:rep_granger}, we have
\[
\eta_n = \mathfrak{F}^{\star}(B)\epsilon_n = \sum_{i=0}^{\infty} \mathfrak{F}^\star_i \epsilon_{n-i}, \quad \text{and} \quad N_n = \mathfrak{F} \cdot \sum_{i=1}^{n} \epsilon_i.
\]

From Theorem B.13 of \citet{johansen_likelihood-based_1995} we get
\begin{align*}
    \sum_{i=2}^T N_{i-1} N_i' &= \sum_{i=2}^T N_{i-1} N_{i-1}' + O_p(T), \\
    \sum_{i=2}^T N_{i-1} \eta_i' &= O_p(T), \\
    \sum_{i=2}^T \eta_{i-1} N_i' &= O_p(T), \\
    T^{-1} \sum_{i=2}^T \eta_{i-1} \eta_i' &= \hat{\Sigma}_\eta(1) + o_p(1),
\end{align*}
where \(\hat{\Sigma}_\eta(1)\) is the sample estimator of the lag-1 autocovariance function of the process \(\eta\).

From these and \eqref{eq:covariance_descomp}, it follows that
\[
S_{XY} = T^{-1} \sum_{i=2}^T N_{i-1} N_{i-1}' + \hat{\Sigma}_\eta(1) + O_p(1).
\]

Similarly, we obtain
\[
S = T^{-1} \sum_{i=1}^T N_i N_i' + \hat{\Sigma}_\eta(0) + O_p(1).
\]

Let \(\tilde{S}_{XY} := T^{-1} S_{XY}\), \(\tilde{S}_{YX} := T^{-1} S_{YX}\), and \(\tilde{S} := T^{-1} S\), then
\[
\tilde{S}_{XY} = T^{-2} \sum_{i=2}^T N_{i-1} N_{i-1}' + o_p(1), \quad \text{and} \quad \tilde{S} = T^{-2} \sum_{i=1}^T N_i N_i' + o_p(1).
\]

This follows from the fact that \(T^{-1} \to 0\) as \(T \to \infty\), hence
\[
T^{-1} \hat{\Sigma}_\eta(i) = o_p(1) \text{ for } i = 0,1, \quad \text{and} \quad T^{-1} O_p(1) = o_p(1) O_p(1) = o_p(1).
\]

Thus, we get
\begin{equation}
\label{eq:equivalence_covariance}
\tilde{S}_{XY} = \tilde{S} + o_p(1), \quad \text{and} \quad \tilde{S}_{YX} = \tilde{S} + o_p(1).
\end{equation}

In \citet{engle_co-integration_1987} it is stated that \(\mathbb{E}[T^{-1}S]\) converges to a non-zero matrix, which implies \(\tilde{S} = O_p(1)\). From this and \eqref{eq:equivalence_covariance}, we have
\[
\tilde{S}_{XY} \tilde{S}_{YX} = \tilde{S}^2 + o_p(1).
\]

On the other hand, note that the first PLS component \(\hat{w}_1\) satisfies the eigenvalue problem
\[
\tilde{S}_{XY} \tilde{S}_{YX} \hat{w}_1 = \lambda_1 \hat{w}_1.
\]

Since \(\tilde{S}_{XY} \tilde{S}_{YX}\) is asymptotically equivalent to \(\tilde{S}^2\), and the eigenvalue problem is invariant under scalar multiplication, we have that in the limit \(\hat{w}_1\) corresponds to the first eigenvector of \(\tilde{S}^2\), i.e., the first principal component.

Hence, we may consider \(P_1 \approx I_m - \hat{w}_1 \hat{w}_1'\), and then the next step of Algorithm \ref{algo:PLS_TS} corresponds to solving the eigenvalue problem:
\[
P_1 \tilde{S}_{XY} \tilde{S}_{YX} P_1 \hat{w}_2 = \lambda_2 \hat{w}_2,
\]
and \(P_1 \tilde{S}_{XY} \tilde{S}_{YX} P_1\) is asymptotically equivalent to \(P_1 \tilde{S}^2 P_1\), so in the limit \(\hat{w}_2\) is the first eigenvector of \(P_1 S^2 P_1\), i.e., the second principal component.

An inductive argument implies that in the \(i\)th iteration, the first eigenvector of \(A_i \tilde{S}_{XY} \tilde{S}_{YX} A_i\) is asymptotically equivalent to the first eigenvector of \(A_i \tilde{S}^2 A_i\), which corresponds to the \(i\)th principal component.

Thus, we have
\[
\hat{\beta}_{\textsc{PLS}} = \hat{\beta}_{\textsc{PCA}} + o_p(1).
\]

Hence,
\[
P_{\beta_\perp} \hat{\beta}_{\textsc{PLS}} = P_{\beta_\perp} \hat{\beta}_{\textsc{PCA}} + o_p(1),
\]
and by Lemma 1 of \citet{harris_principal_1997} we have \(P_{\beta_\perp} \hat{\beta}_{\textsc{PCA}} = O_p(T^{-1})\), so that
\[
P_{\beta_\perp} \hat{\beta}_{\textsc{PLS}} = O_p(T^{-1}),
\]
which concludes the proof. \qed

\begin{figure}[!ht]
    \centering
    \includegraphics[width=0.7\linewidth]{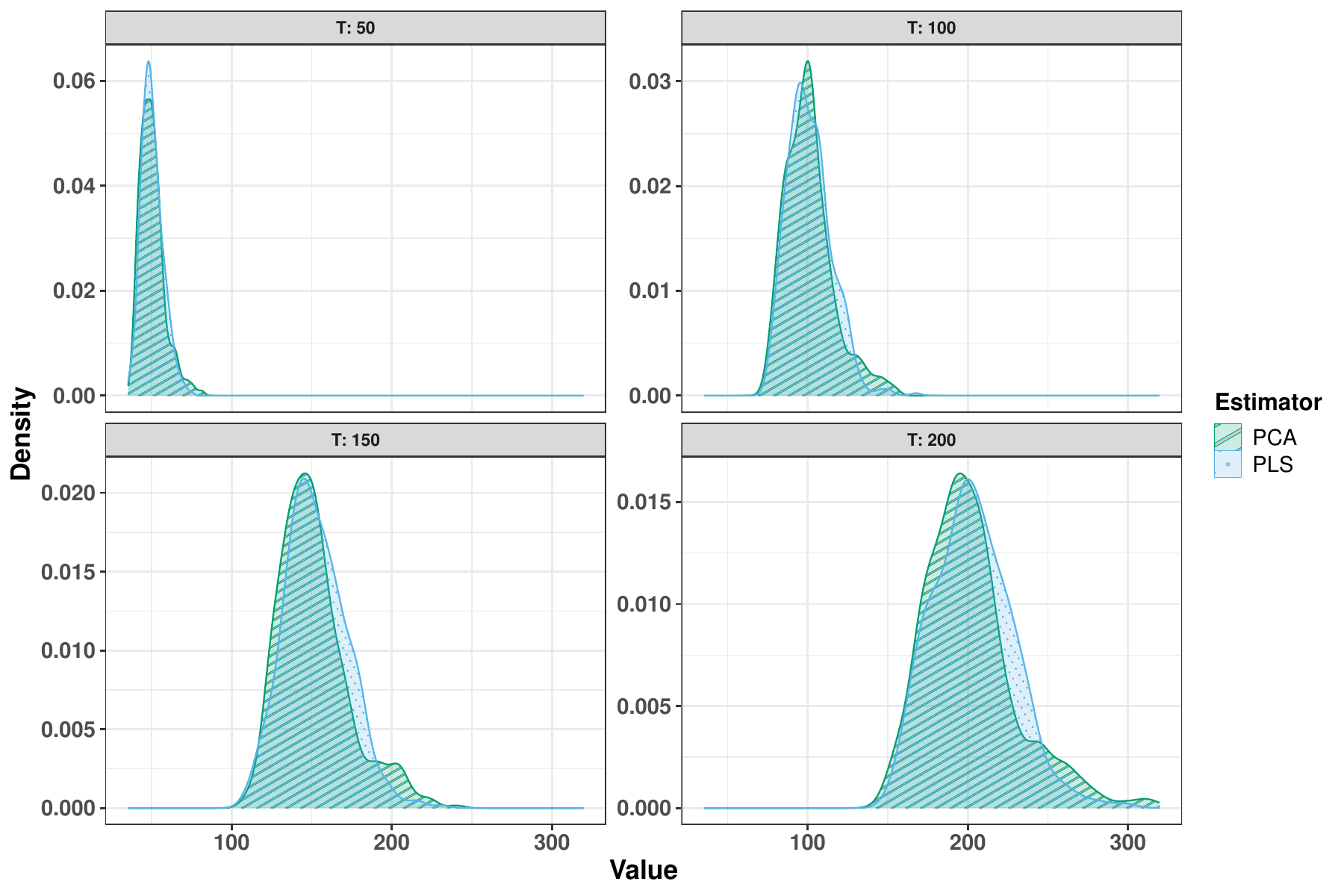}
    \caption{Visualisation of the distribution of $T\norm{P_{\beta_\perp}\hat{\beta}}$ under Assumption \ref{ass:cointegration} for PCA and PLS methods considering $S=500$, $m=11$ and $r=9$.}
    \label{fig:asymptotics}
\end{figure}

\section{Software}
\label{append:software}
Link to a GitHub repository with all the source code for reproducing the tables and plots of the paper: \href{https://github.com/robervz22/ts_stable_space}{ts\_stable\_space}.

\end{document}